\documentclass[onecolumn,11pt]{article}

\usepackage[top=1in, bottom=1in, left=1in, right=1in]{geometry}
\usepackage{amsmath,amsfonts,amscd,amssymb}
\usepackage[utf8]{inputenc} 
\usepackage[T1]{fontenc}    
\usepackage{hyperref}       
\usepackage{url}            
\usepackage{booktabs}       
\usepackage{amsfonts}       
\usepackage{nicefrac}       
\usepackage{microtype}      
\usepackage{indentfirst}
\usepackage{hyphenat}
\usepackage{amsmath}
\usepackage{amssymb}
\usepackage{mathrsfs} 
\usepackage{mathtools}

\usepackage{graphicx}
\usepackage{caption}
\usepackage{float}
\usepackage[abs]{overpic}
\usepackage{subfig}
\usepackage{epstopdf}
\usepackage{rotating}
\usepackage{multirow}
\usepackage{wrapfig}

\usepackage{enumerate}
\usepackage[numbers,sort&compress]{natbib}

\usepackage[noend]{algpseudocode}
\usepackage[utf8]{inputenc}
\usepackage{fontenc}
\usepackage{lipsum}
\usepackage[ruled,vlined]{algorithm2e}

\usepackage[bottom,flushmargin,hang,multiple]{footmisc}
\usepackage{lipsum}
\usepackage{cancel}
\usepackage{url}
\usepackage{color}
\usepackage{mathtools}
\usepackage{subeqnarray}
\usepackage{setspace}
\usepackage{palatino} 
\usepackage{appendix}
\usepackage{soul} 

\definecolor{header1}{cmyk}{0,0,0,1}

\newcommand{\bn}{\mathbf{n}}
\newcommand{\bhn}{\mathbf{\hat{n}}}

\newcommand{\bx}{\mathbf{x}}

\newcommand{\by}{\mathbf{y}}

\newcommand{\bX}{\mathbf{X}}
\newcommand{\bhX}{\mathbf{\hat{X}}}
\newcommand{\bhx}{\mathbf{\hat{x}}}
\newcommand{\bdhX}{\mathbf{\dot{\hat{X}}}}
\newcommand{\bN}{\mathbf{N}}
\newcommand{\bhN}{\mathbf{\hat{N}}}
\newcommand{\bY}{\mathbf{Y}}
\newcommand{\bF}{\mathbf{F}}
\newcommand{\bhF}{\mathbf{\hat{F}}}

\newcommand{\bsf}{\boldsymbol{f}}
\newcommand{\bsg}{\boldsymbol{g}}
\newcommand{\bshf}{\boldsymbol{\hat{f}}}

\newcommand{\bXi}{\boldsymbol{\Xi}}
\newcommand{\bxi}{\boldsymbol{\xi}}

\newcommand{\bTheta}{\boldsymbol{\Theta}}

\newcommand\blfootnote[1]{%
  \begingroup
  \renewcommand\thefootnote{}\footnote{#1}%
  \addtocounter{footnote}{-1}%
  \endgroup
}

\title{\vspace{-.4in}{\fontsize{15}{15}\selectfont \textbf{
Automatic Differentiation to Simultaneously Identify Nonlinear Dynamics and Extract Noise Probability Distributions from Data
}}}
\author{\normalsize{Kadierdan Kaheman$^{1*}$, Steven L. Brunton$^1$, J. Nathan Kutz$^2$}\\
\footnotesize{$^1$ Department of Mechanical Engineering, University of Washington, Seattle, WA 98195, United States}\\
\footnotesize{$^2$ Department of Applied Mathematics, University of Washington, Seattle, WA 98195, United States\vspace{-.2in}}
}
\date{}

\begin{document}
\maketitle
\blfootnote{$^*$ Corresponding author (kadierk@uw.edu); Code available at github.com/dynamicslab/modified-SINDy.}

\vspace{-.2in}
\begin{abstract}
The {\em sparse identification of nonlinear dynamics} (SINDy) is a regression framework for the discovery of parsimonious dynamic models and governing equations from time-series data.  
As with all system identification methods, noisy measurements compromise the accuracy and robustness of the model discovery procedure.
In this work we develop a variant of the SINDy algorithm that integrates automatic differentiation and recent time-stepping constrained motivated by Rudy et al.~\cite{rudy2019deep} for simultaneously (i) denoising the data, (ii) learning and parametrizing the noise probability distribution, and (iii) identifying the underlying parsimonious dynamical system responsible for generating the time-series data.
Thus within an integrated optimization framework, noise can be separated from signal, resulting in an architecture that is approximately twice as robust to noise as state-of-the-art methods, handling as much as 40\% noise on a given time-series signal and explicitly parametrizing the noise probability distribution.  
We demonstrate this approach on several numerical examples, from Lotka-Volterra models to the spatio-temporal Lorenz 96 model.  
Further, we show the method can identify a diversity of probability distributions including Gaussian, uniform, Gamma, and Rayleigh.
\end{abstract}

\section{Introduction}
\label{sec1}

The data-driven discovery of governing equations is an emerging field within the machine learning and artificial intelligence communities.  
From {\em neural networks} (NN) to traditional model regression techniques, a diversity of methods are emerging that transform time-series data (or spatio-temporal data) into representations of governing equations of motion~\cite{Brunton2019book,Nelles2013book,ljung2010arc,Schmid2010jfm,Kutz2016book,Budivsic2012chaos,Mezic2013arfm,Williams2015jnls,klus2017data,Akaike1969annals,Billings2013book,yang2018physics,Wehmeyer2018jcp,Mardt2018natcomm,vlachas2018data,pathak2018model,lu2019deepxde,Raissi2019jcp,Champion2019pnas,raissi2020science,Raissi2017arxiva,Raissi2017arxiv,Giannakis2012pnas,Yair2017pnas,Bongard2007pnas,Schmidt2009science,Daniels2015naturecomm,Schaeffer2017prsa,Brunton2016SINDy,Rudy2017PDE-find,rudy2019deep,yao2007modeling}. 
The interpretability and generalizability of these discovered equations of motion are critical for understanding, designing, and controlling complex systems.  
As such, the {\em sparse identification of nonlinear dynamics} (SINDy)~\cite{Brunton2016SINDy} framework provides a compelling regression framework since discovered models are interpretable and parsimonious by design.  
As with all system identification algorithms, noisy measurements compromise the accuracy and robustness of the model discovery procedure.
Moreover, many optimization frameworks rely explicitly on the assumption of Gaussian noise, which is rarely true in the real world.  
Recently, Rudy et al.~\cite{rudy2019deep} developed a novel optimization framework for separating signal and noise from noisy time-series data by identifying a deep NN model for the signal from numerical time-stepping constraints such as a Runge-Kutta.
In this work, we build on this framework and leverage automatic differentiation ~\cite{baydin2017automatic} in the optimization procedure
to simultaneously denoise data and identify sparse nonlinear models via SINDy. %
This new architecture yields significant improvements in model discovery, including superior separation of the signal from noise while simultaneously characterizing the noise distribution. 


SINDy has emerged as a flexible and promising architecture for model discovery due to its inherent parsimonious representation of dynamics.  
The SINDy framework relies on sparse regression on a library of candidate model terms to select the fewest terms required to describe the observed dynamics~\cite{Brunton2016SINDy}.  
Specifically, SINDy is formulated as an over-determined linear system of equations $\mathbf{A}\bxi={\bf b}$, where sparsity of the solution is promoted by the $\ell_0$-norm $\|{\bxi}\|_0$.  
Thus sparsity is a proxy for parsimony, interpretability, and generalizability.  
Measurement noise, however, is always present, and it corrupts the ability of the SINDy regression framework, and indeed any other model discovery paradigm, to accurately extract governing models.  

There are many variants of sparse regression, all of which typically attempt to approximate the solution to an NP-hard, $\ell_0$-norm penalized regression. Sparsity-promoting methods like the LASSO~\cite{su2017false,Tibshirani1994} use the $\ell_1$-norm as a proxy for sparsity since tractable computations can be performed. 
The iterative least-squares thresholding technique of the SINDy algorithm promotes sparsity through a sequential procedure.  
Recently, Zhang and Schaeffer~\cite{zhang2019convergence} have provided several rigorous theoretical guarantees on the convergence of the SINDy algorithm. 
Specifically, they proved that the algorithm approximates local minimizers of an unconstrained $\ell_0$-penalized least-squares problem, which allows them to provide sufficient conditions for general convergence, the rate of convergence, and conditions for one-step recovery. Using a relaxed formulation, Champion et al.~\cite{Champion2019SINDySR3} show how the SINDy regression framework can accommodate additional structure, robustness to outliers, and nonlinear parameter estimation using the {\em sparse relaxed regularized regression} (SR3) formulation~\cite{ZhengSR3}.  SINDy results in interpretable models, and it has been widely applied in many scientific disciplines~\cite{Sorokina2016oe,JC2018ConGalerkin,Dam2017pf,Loiseau2018jfm,Hoffmann2018arxiv,loiseau2020data,el2018sparse,narasingam2018data,de2019discovery,KK2019Discrepancy,Thaler2019jcp,lai2019sparse,Deng2020JFM,schmelzer2020discovery,pan2020sparsity,beetham2020formulating}. Moreover, it has been extended to incorporate control~\cite{Brunton2016SINDyc,Kaiser2018prsa}, rational or implicit dynamics~\cite{Mangan2016ImSINDY,kaheman2020sindypi,Zhang2018}, partial differential equations~\cite{Rudy2017PDE-find,messenger2020weakpde}, parametric model dependencies~\cite{Rudy2019siads}, discrepancy models~\cite{KK2019Discrepancy,de2019discovery}, multiscale physics~\cite{champion2019discovery}, stochastic dynamics~\cite{boninsegna2018sparse}, constrained physics~\cite{Loiseau2018jfm}, among many other innovations~\cite{JC2018ConGalerkin,Schaeffer2017prsa,Mangan2017ModelSelection,tran2017exact,Schaeffer2017pre,schaeffer2018extracting,wu2018numerical,boninsegna2018sparse,mangan2019model,Gelss2019mindy,Champion2019SINDySR3,goessmann2020tensor,Reinbold2020pre,messenger2020weaksindy}. The \texttt{PySINDy} Python package executes many of these variants~\cite{de2020pysindy}. 

Despite its flexibility, modularity, and extensibility, SINDy and its variants typically rely on approximating time-derivative of the measured time-series data. Computing derivatives of noisy measurement data is known to be a challenging problem, with many algorithmic innovations and mathematical architectures developed to produce accurate derivative approximations~\cite{floris}. These methods include finite-differences, spectral methods~\cite{kutz2013data}, spline smoothing, filtering procedures, polynomial fitting, low-rank projection, and total variations, to highlight some of the diverse techniques employed for this critical task of scientific computing. This task is made even more difficult, depending upon the noise statistics. Gaussian noise is often easier to learn and characterize than noise distributions that have non-zero means and are not symmetric. Ultimately, there is a need for methods that are robust to noisy measurements and diverse probability distributions. 

Recent innovations in automatic differentiation have enabled the solution of an optimization problem directly related to the computation of the required derivatives~\cite{rudy2019deep}. Since its inception, automatic differentiation has been widely used in the machine learning community to enable complicated optimization problems without manually computing Jacobians~\cite{abadi2016tensorflow,baydin2017automatic,rackauckas2017differentialequations,van2018automatic,chen2018neural,rudy2019smoothing,rudy2019deep,both2019deepmod,rackauckas2020universal,lange2020fourier}. 
More recently, this approach has been used with NNs to separate a signal from noise and model the signal when a model is unknown~\cite{rudy2019deep}, and to improve Kalman smoothing when the governing equations are known~\cite{rudy2019smoothing}. The success of these algorithms suggest that they could be leveraged for noise signal separation in the SINDy framework. In this work, we extend this simultaneous de-noising and discovery approach to SINDy. Specifically, automatic differentiation enables differentiation with respect to the functions in the SINDy library, thus circumventing a direct differentiation of the noisy time-series data. The modified SINDy algorithm is more robust to noise and further allows for an explicit characterization (discovery) of the underlying probability distribution of the noise, something that current state-of-the-art methods cannot do and is a unique feature of our method.  

In Sec.~2, we illustrate the modified SINDy algorithm. In Sec.~3, we show the comparison between modified SINDy and noise signal separation approach based on the NN proposed by Rudy et al.~\cite{rudy2019deep}. In Sec.~4, we show the use of modified SINDy on various numerical examples. We also show how modified SINDy can be used to extract the noise distribution information and how it can be used in the discrepancy modeling framework. In Sec.~5, we show our conclusions and possible future improvements.

\section{Methods}
\label{secMethod}
In what follows, we introduce the basic mathematical architecture behind the SINDy algorithm, demonstrating explicitly its sensitivity to noisy measurements. This guides our introduction of the modified SINDy for simultaneously learning the system model and denoising the signal.

\subsection{Sparse Identification of Nonlinear Dynamics}
\label{secMethod:SINDy}
The SINDy algorithm~\cite{Brunton2016SINDy} provides a principled, data-driven discovery method for nonlinear dynamics of the form
\begin{equation}
    \frac{d}{dt}\bx(t)=\bsf(\bx(t)),
    \label{eq:dx=f(x)}
\end{equation}
where $\bx(t)=[x_1(t),x_2(t),\cdots,x_n(t)]\in \mathbb{R}^{1\times n}$ is system states represented as a row vector.
%
%
SINDy posits a set of candidate functions that would characterize the right hand side of the governing equations.  Candidate model terms form the library $\bTheta(\bX)=[{\theta}_1(\bX),{\theta}_2(\bX),\cdots,{\theta}_p(\bX)]\in \mathbb{R}^{m\times p}$ of potential right hand side terms, where $\bX=[\bx(t_1);\bx(t_2);\cdots;\bx(t_m)]\in \mathbb{R}^{m\times n}$ is formed by $m$ row vectors. This then allows for the formulation of a regression problem to select only the few candidate terms necessary to describe the dynamics:
\begin{equation}
    \begin{gathered}
        \arg \min_{\bXi} \| \dot{\bX} - \bTheta(\bX) \bXi \|_2 + \lambda \|\bXi\|_0,
    \end{gathered}
    \label{eq:dX=ThetaXi}
\end{equation}
where the matrix $\boldsymbol{\Xi}=[{\boldsymbol{\xi}_{1}},\boldsymbol{\xi}_{2},\cdots,{\boldsymbol{\xi}_{n}}]\in \mathbb{R}^{p\times n}$ is comprised of the sparse vectors ${\boldsymbol{\xi}_{i}}\in \mathbb{R}^{p\times 1}$ that select candidate model terms.  The amount of sparsity promotion is controlled by the parameter $\lambda$, which determines the penalization by the $\ell_0$-norm. The $\theta_i(\bX)$ can be any candidate function that may describe the system dynamics $\bsf(\bx(t))$ such as  trigonometric functions ${\theta}_i(\bX)=\cos (\bX)$ or polynomial functions ${\theta}_i(\bX)=\bX^3$, for example. By solving Eq.~\eqref{eq:dX=ThetaXi}, we can identify a model of system dynamics
\begin{align}
    \label{eq:PerfectDataSINDy}
    \frac{d}{dt}\bx(t)=\bsf(\bx(t))\approx \bTheta(\bx(t))\bXi.
\end{align} 
Many different optimization techniques can be used to obtain the sparse coefficients $\bXi$, such as sequentially thresholded least squares~(STLSQ)~\cite{Brunton2016SINDy,zhang2019convergence}, LASSO~\cite{Tibshirani1994}, sparse relaxed regularized regression~(SR3)~\cite{ZhengSR3,Champion2019SINDySR3}, stepwise sparse regression~(SSR)~\cite{Boninsegna2018SSR}, and Bayesian approaches~\cite{Zhang2018,Pan2016BayesianSINDy}. 

In practice, noise-free measurements of $\bx(t)$ are not available, and only the full state noisy measurement
\begin{equation}
    \by(t)=\bx(t)+\bn(t),
    \label{eq:y}
\end{equation}
is provided to SINDy from sensors, where $\by(t)=[y_1(t),y_2(t),\cdots,y_n(t)]\in \mathbb{R}^{1\times n}$ is noisy measurement and $\bn(t)=[n_1(t),n_2(t),\cdots,n_n(t)]\in \mathbb{R}^{1\times n}$ is the noise added to true state. Thus,  Eq.~\eqref{eq:dX=ThetaXi} then becomes
\begin{equation}
    \dot{\bY}=\dot{\bX}+\dot{\bN}=\bTheta(\bY)\bXi=\bTheta(\bX+\bN)\bXi,
    \label{eq:dY=dTheta(Y)}
\end{equation}
where $\bY=[\by(t_1);\by(t_2);\cdots;\by(t_m)]\in \mathbb{R}^{m\times n}$ is noisy measurement matrix formed by $m$ row vectors measurement of size $1\times n$ and $\bN=[\bn(t_1);\bn(t_2);\ \cdots;\bn(t_m)]\in \mathbb{R}^{m\times n}$ is noise matrix also formed by $m$ row vector of size $1\times n$. From Eq.~\eqref{eq:dY=dTheta(Y)}, note that the solution $\bXi$ is no longer the same $\bXi$ shown in Eq.~\eqref{eq:dX=ThetaXi} due to the presence of noise. Moreover, the noise will be magnified when approximating the derivatives $\dot{\bX}$ by a factor of $\mathcal{O}(1/dt)$~\cite{Rudy2017PDE-find}, and it will non-linearly corrupt the library matrix $\bTheta$. Extensive research has been done to improve the robustness of the SINDy framework. The integral formulation~\cite{Schaeffer2017pre} and weak formulation~\cite{Reinbold2020pre,messenger2020weaksindy,messenger2020weakpde} improved the regression robustness by avoiding taking derivative of noisy data. Other approaches, such as subsampling~\cite{zhang2019robust}, increased the noise robustness of the SINDy framework by doing regression on the subsampled measurement that has less noise.  Corrupt data can also be handled with methods from robust statistics~\cite{tran2017exact,Champion2019SINDySR3}. In the next section, we introduce an alternative approach that simultaneously learns the noise $\bN$ while using the denoised data to perform model identification.

\subsection{Simultaneously Denoising and Learning System Model}
\label{secMethod:modified SINDy}
\begin{figure}[t]
    \centering
    \includegraphics[width=\textwidth]{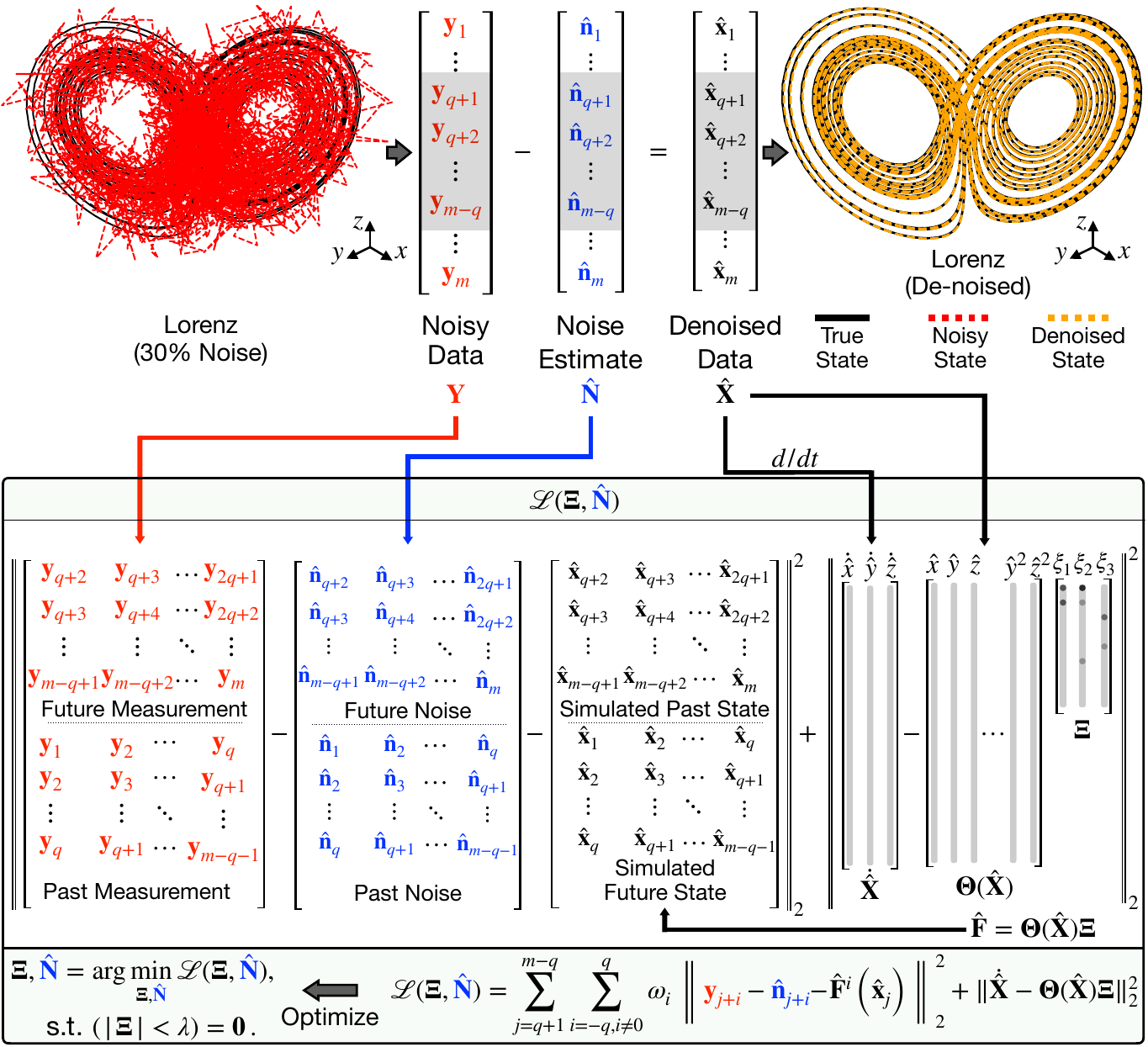}
    \caption{This figure illustrates modified SINDy algorithm. The goal is to learn the system model $\bTheta(\bX)\bXi$ and noise $\mathbf{N}$. The noise is subtracted from the measurement to obtain the clean data. To achieve this, the estimated noise $\bhN$ is set as an optimization parameter and the cost function $\mathcal{L}(\bXi,\bhN)$ is minimized. This optimization is performed several times, and the small values of $|\bXi|$ is enforced to be zero for the remainder of the optimization process. }
    \label{FigMethod}
\end{figure}
To improve the noise robustness of the SINDy regression, we determine the estimated noise $\bhn(t)\in \mathbb{R}^{1\times n}$ as a hyper-parameter and formulate $\bhN=[\bhn(t_1);\bhn(t_2);\cdots;\bhn(t_m)]\in\mathbb{R}^{m\times n}$ in order to optimize the difference between the estimated derivative and system's vector field such that
\begin{equation}
    e_d=\|\bdhX-\bTheta(\bhX)\bXi\|_2^2,
    \label{eq:ed}
\end{equation}
where $\bhX=\bY-\bhN$ is formed by $m$ estimated true states $\bhx(t)\in \mathbb{R}^{1\times n}$, and $e_d$ is the derivative approximation error. Note that $\bhX=[\bhx(t_1);\bhx(t_2);\cdots;\bhx(t_m)]\in\mathbb{R}^{m\times n}$. When $\bhN=\bN$, the effect of noise can be eliminated, and the accuracy of SINDy will be significantly improved. However, there exist many trivial solutions for minimizing the Eq.~\eqref{eq:ed} with two uncorrelated optimization parameters $\bhN$ and $\bXi$. Thus, an additional constraint is needed to regularize Eq.~\eqref{eq:ed}.

The additional constraint proposed here uses the estimated vector field of the system model similar to the one proposed by Rudy et al.~\cite{rudy2019deep}. Equation~\eqref{eq:PerfectDataSINDy} gives the estimate $\bTheta(\bx(t))\bXi$ of the true vector field $f(\bx (t))$.  Integrating over a segment of time $t_j$ to $t_{j+1}$ gives the integrated vector field, or flow map,
\begin{equation}
    \label{eq:PerfectDataSimulateOneStep}
    \bx(j+1)=\bF(\bx(j))=\bx(j)+\int_{t_j}^{t_{j+1}} \bTheta(\bx(\tau))\bXi\ d\tau.
\end{equation}
This can be generalized to integrate the system either forward or backward in time $q$ steps.  This gives 
\begin{equation}
    \label{eq:PerfectDataSimulateQSteps}
    \bx(j+q)=\bF^q(\bx(j))=\bx(j)+\int_{t_j}^{t_{j+q}} \bTheta(\bx(\tau))\bXi\ d\tau.
\end{equation}
To obtain the $\bx(j+q)$ in Eq.~\eqref{eq:PerfectDataSimulateQSteps}, a numerical simulation scheme such as Runge–Kutta can be used. In what follows, we employ a 4th-order Runge-Kutta method to simulate the dynamics forward/backward in time $q$-steps. Similar to Eq.~\eqref{eq:PerfectDataSimulateQSteps}, when the noisy measurement data $\by$ is given, the estimated state $\bhx=\by-\bhn$ satisfies
\begin{equation}
    \label{eq:NoisyDataSimulateQSteps}
    \by(j+q)-\bhn(j+q)=\bhx(j+q)=\bhF^q(\bhx(j))=\bhx(j)+\int_{t_j}^{t_{j+q}} \bTheta(\bhx(\tau))\bXi\ d\tau,
\end{equation}
when $\bhn=\bn$ and the exact value of $\bXi$ is known. Thus, by minimizing
\begin{equation}
    \label{eq:minSimbshx(i+q)}
    e_{s,j}=\sum_{i=-q,i\neq0}^{q}\omega_i\|\by(j+i)-\bhn(j+i)-\bhF^i(\bhx(j))\|_2^2,
\end{equation}
the optimization parameters $\bhN$ and $\bXi$ are coupled, resulting in additional structural constraint of the model. The parameter $\omega_i$ is used to account for the numerical error and is set to $\omega_i=c^{|i|-1}$, where $0<c\leq1$ is a constant~(throughout this paper, we use $c=0.9$). The use of $\omega$ suggests that the simulation error too far ahead in the future, or too far backward in the past, should be penalized less due to the error of the numerical simulation scheme. The error incurred by simulating the vector filed forward/backward on the entire trajectory can be written as
\begin{equation}
    \label{eq:minSimAllPoint}
    e_{s}=\sum_{j=q+1}^{m-q}e_{s,j}=\sum_{j=q+1}^{m-q}\sum_{i=-q,i\neq0}^{q}\omega_i\|\by(j+i)-\bhn(j+i)-\bhF^i(\bhx(j))\|_2^2.
\end{equation}
Using subscripts to represent the time step, the final cost function is then
\begin{equation}
    \label{eq:costfunc}
    \mathcal{L}(\bXi,\bhN)=e_s+e_d=\sum_{j=q+1}^{m-q} \sum_{i=-q,i\neq0}^{q} \omega_i\|\by_{j+i}-\bhn_{j+i}-\bhF^{i}\left(\bhx_{j}\right)\|_2^2+\|\bdhX-\bTheta(\bhX)\bXi\|_2^2,
\end{equation}
which is the summation of the derivative approximation error $e_d$ and simulation error $e_s$. The optimization problem to simultaneously denoise and learn the system model can then be written as
\begin{equation}
    \label{eq:optProb}
    \begin{gathered}
        \bXi, \bhN  =\arg \min_{\bXi, \bhN}\mathcal{L}(\bXi, \bhN),\\
        \textrm{s.t.} \quad (|\bXi|<\lambda) =\mathbf{0}.
    \end{gathered}
\end{equation}

The global optimal solution for Eq.~\eqref{eq:optProb} needs to satisfy $\bhN=\bN$ and $\bsf(\bx)=\bTheta(\bx)\bXi$. To solve for Eq.~\eqref{eq:optProb}, it is necessary to calculate the Jacobian $\partial \mathcal{L}/\partial \bhN$ and $\partial \mathcal{L}/\partial \bXi$, which is a difficult task to do analytically or computationally. However, recent automatic differentiation packages such as Tensorflow~\cite{abadi2016tensorflow} and Julia Flux~\cite{innes2018} make it possible to directly extract the gradients of $\mathcal{L}$ with respect to $\bhN$ and $\bXi$.  This allows us to solve the optimization problem in Eq.~\eqref{eq:optProb} easily using gradient descent method such as Adam~\cite{kingma2014adam}. Throughout this paper, we use the Tensorflow 2.0 and Adam optimizer to solve the Eq.~\eqref{eq:optProb}. Moreover, to enforce the sparsity of the identified model, a thresholding approach~\cite{Brunton2016SINDy} is used and the Eq.~\eqref{eq:optProb} is solved for $N_{loop}$ times. Each iteration uses the previous iteration's optimization result $\bhN$ as the initial guess of the new iteration. The values of $\bhN$ is also used to calculate the estimated state $\bhX$, which is used to calculate the new estimated values of the selection parameter $\bXi$. Furthermore, if the elements in $|\bXi|$ are smaller than a threshold $\lambda$ at the end of an optimization loop, those elements will be constrained to zero for the remainder of the optimization process. Figure~\ref{FigMethod} illustrates this process, and Appendix.~\ref{Appendix:Algorithm} shows the detailed algorithm for simultaneous denoising and sparse model identification. Some guidance on the selection of the hyper-parameters $\lambda$, $q$, and $N_{loop}$ is given in Appendices.~\ref{Appendix:EffectOfLambda},~\ref{Appendix:EffectOfPreStep}, and~\ref{Appendix:EffectOfNloop}.
\section{Performance Comparison with Neural Network Denoising Approach}
The advocated optimization framework of modified SINDy is compared with a NN denosing approach by Rudy et al.~\cite{rudy2019deep}. Additionally, the robustness to noise and the amount of data is considered.

\subsection{Performance Criteria}
For ease of comparison, we use the same performance criteria developed by Rudy et al.~\cite{rudy2019deep}.  Specifically, these are the vector field error $E_{\bsf}$, the noise identification error $E_{\bN}$, and the prediction error $E_{\bF}$. The vector filed error is
\begin{equation}
    \label{eq:Ef}
    E_{\bsf}=\frac{\sum_{i=1}^{m}\left\|\bsf\left(\mathbf{x}_{i}\right)-\bshf\left(\mathbf{x}_{i}\right)\right\|_{2}^{2}}{\sum_{i=1}^{m}\left\|\bsf\left(\mathbf{x}_{i}\right)\right\|_{2}^{2}},
\end{equation}
which calculates the relative squared $\ell_2$ error between the true vector filed and identified vector field $\bshf$. The noise identification error is 
\begin{equation}
    \label{eq:En}
    E_{\bN}=\frac{1}{m} \sum_{i=1}^{m}\left\|\bn_{i}-\bhn_{i}\right\|_{2}^{2},
\end{equation}
which is the mean $\ell_2$ difference between the true noise $\bN$ and identified noise $\bhN$. The prediction error is 
\begin{equation}
    \label{eq:EF}
    E_{\bF}=\frac{1}{\|\mathbf{X}\|_{F}^{2}} \sum_{i=1}^{m-1}\left\|\mathbf{x}_{i}-\bhF^{i}\left(\mathbf{x}_{1}\right)\right\|_{2}^{2},
\end{equation}
and it calculates the difference between forward simulation trajectory and true trajectory. For comparison of modified SINDy and recently published Weak-SINDy~\cite{messenger2020weaksindy}, as shown in Appendix.~\ref{Appendix:WeakSINDy}, two more performance criteria are used. The first one is the normalized parameter error
\begin{equation}
    \label{eq:Ep}
    E_{\boldsymbol{p}}=\frac{\|\bXi-\hat{\bXi}\|_2}{\|\bXi\|_2},
\end{equation}
which reflects how much the identified parameters $\hat{\bXi}$ is off from the true parameters $\bXi$. The other one is the success rate, which describes the percentage of identifying the model's correct structure in multiple trials.

\subsection{Robustness to Noise}
\label{sec:CompareNoiseNNandModifiedSINDy}
The Lorenz attractor is used as an example to test the noise robustness of the the approach. The model of the chaotic Lorenz is
\begin{equation}
    \label{eq:lorenz}
    \begin{aligned}
        \dot{x}&=\sigma(y-x),\\
        \dot{y}&=x(\rho-z)-y,\\
        \dot{z}&=x y-\beta z,
    \end{aligned}
\end{equation}
where $\sigma=10$, $\rho=28$, and $\beta=8/3$. The Lorenz attractor is simulated with initial condition $x_0=[5,5,25]$, $T=25$, and $dt=0.01$. The prediction step is chosen as $q=3$ for both approaches compared and $N_{loop}=6$ for our proposed method. Unless otherwise noted, Adam optimizer is used to optimize the problem with maximum iteration set to $5000$ for modified SINDy and $30000$ for NN approach~\cite{rudy2019deep}. Different magnitudes of Gaussian noise are added to generate the noisy training data. The noise level is defined as
\begin{equation}
    \label{eq:NoiseLevel}
    \text{Noise Level}~(\%)=\frac{\mbox{var}(\text{Noise)}}{\mbox{var}(\text{Signal})}\times 100\%.
\end{equation}
\begin{figure}[t]
    \centering
    \includegraphics[width=1\textwidth]{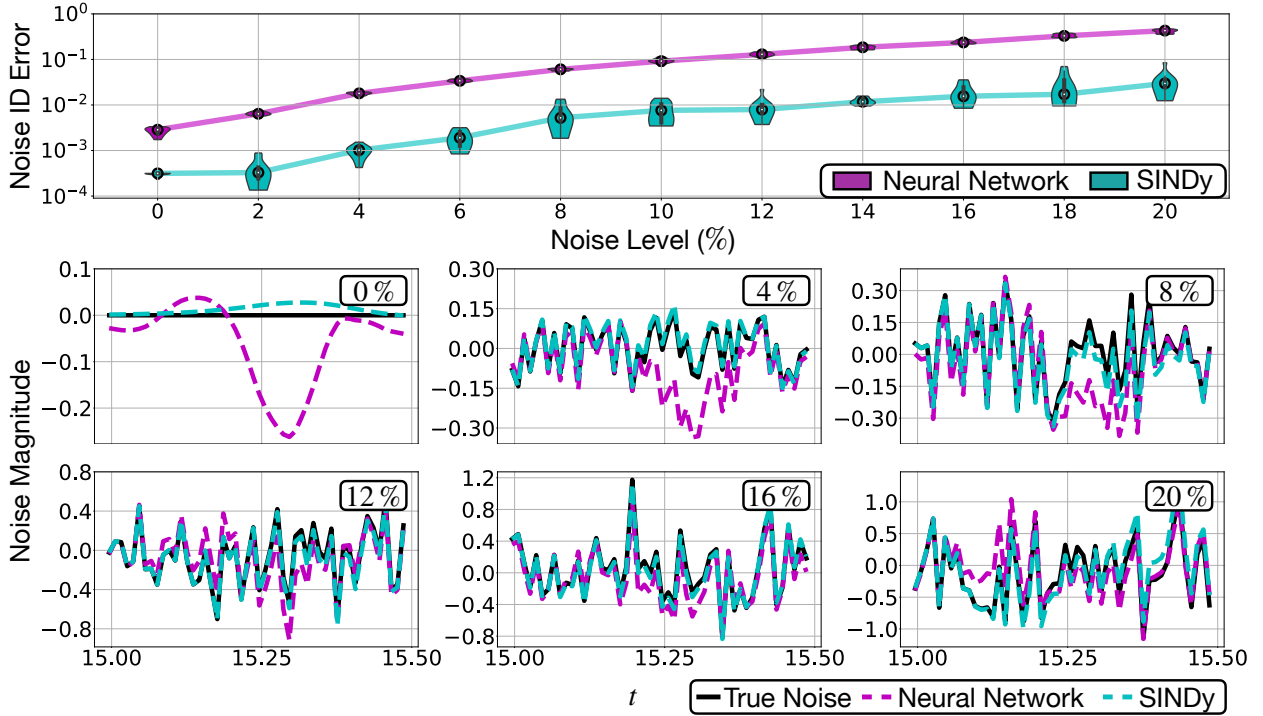}
    \caption{Top: Noise identification error of modified SINDy~(labeled as SINDy) and NN denoising approach by Rudy et al.~\cite{rudy2019deep}. The black circle represents the median of ten runs while the violin shape represents the distribution of error. The modified SINDy approach shows better noise identification error. Bottom: Comparison between the average noise applied to the Lorenz system and the noise identified by the two approaches. As shown on the left, both approaches can not produce the correct zero noise result when no noise is applied, which happens since there is a tiny difference between the learned dynamics and true dynamics.}
    \label{FigCompareNoise1}
\end{figure}

For each noise level, $10$ different sets of noisy data are generated and used as data for both approaches. The NN approach~\cite{rudy2019deep} uses the same set up as~\cite{rudy2019deep}, with $3$ hidden layers, and each layer having $64$ neurons.  Moreover, the regularization parameter is chosen as $10^{-8}$, and the penalty for $\bhN$ is chosen as $10^{-5}$. Unless otherwise noted, we use the same set up for all the NNs in this paper. For modified SINDy, the library is constructed with terms up to second order~(not including the constant term). Moreover, the value of the sparsity parameter $\lambda$ varies based on the noise added. For most of the case, $\lambda=0.1$. A Tikhonov regularization approach is used to pre-smooth the noisy data as in~\cite{rudy2019deep}, although we have found that pre-smoothing does not affect the results appreciably when using zero-mean noise.

\begin{figure}[t]
    \centering
    \includegraphics[width=1\textwidth]{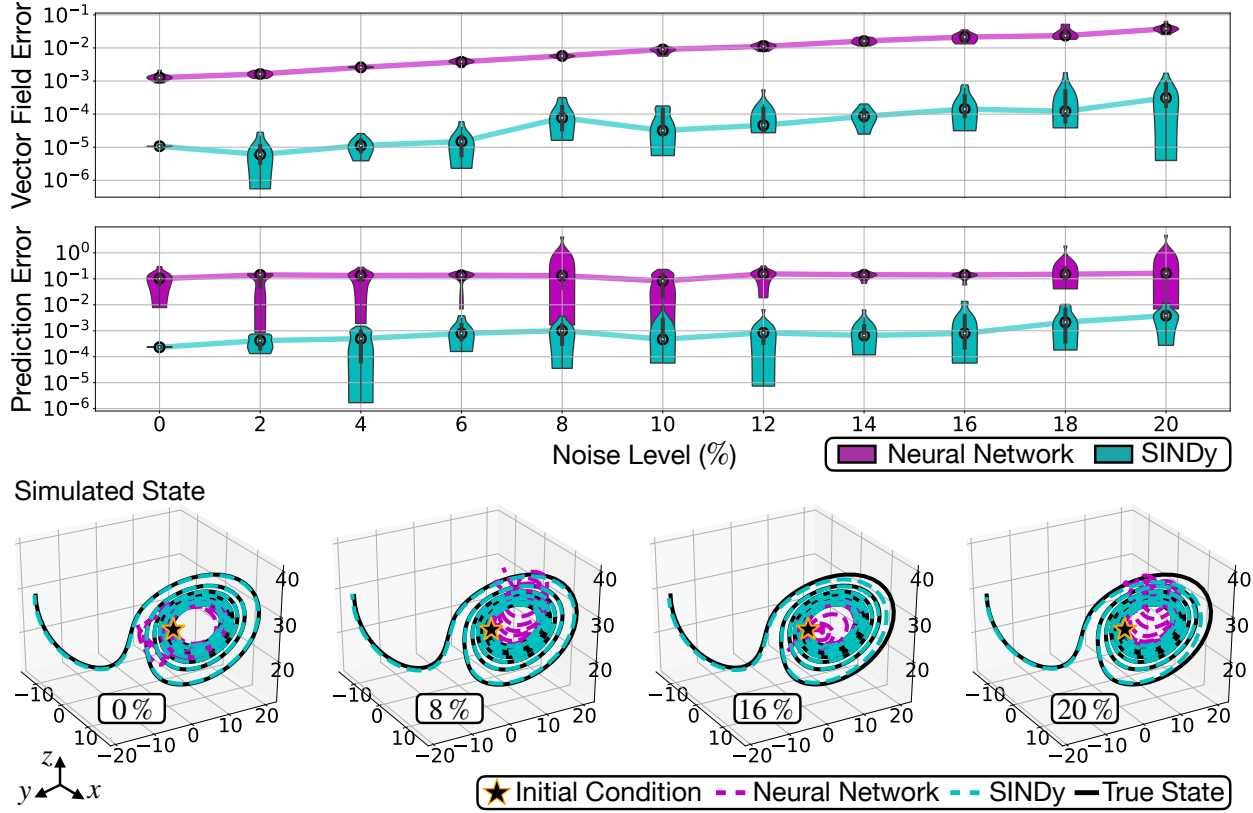}
    \caption{ Top: The vector field error and prediction error of modified SINDy~(labeled as SINDy) and NN denoising approach by Rudy et al.~\cite{rudy2019deep} is shown. The black dot is the median of the $10$ runs, and the violin shape represents the distribution of the error. Bottom: The simulated trajectory is shown with the initial condition chosen as $x_0=[5,5,25]$. }
    \label{FigCompareNoise2}
\end{figure}

Fig.~\ref{FigCompareNoise1} show the noise identification error of the NN approach~\cite{rudy2019deep} and the modified SINDy approach. The vector field error and short term prediction error can be seen in Fig.~\ref{FigCompareNoise2}. For all the noise levels, modified SINDy correctly identified the Lorenz model. To calculate the prediction error, the identified model is simulated $6$ seconds forward in time, with $dt=0.01$, for both modified SINDy and NN denoising approach~\cite{rudy2019deep}. Fig.~\ref{FigCompareNoise2} suggests that modified SINDy identified model has better performance when simulated forward in time. Appendix.~\ref{Appendix:SINDy} shows noise robustness comparison between the modified SINDy and original SINDy~\cite{Brunton2016SINDy}. In general, the modified SINDy is about $2$ times more robust than orignal SINDy~\cite{Brunton2016SINDy}. A comparison between modified SINDy and the recently developed Weak-SINDy approach~\cite{messenger2020weaksindy} is presented in Appendix.~\ref{Appendix:WeakSINDy}.

\subsection{Robustness to Data Length}
\begin{figure}[t]
    \centering
    \includegraphics[width=1\textwidth]{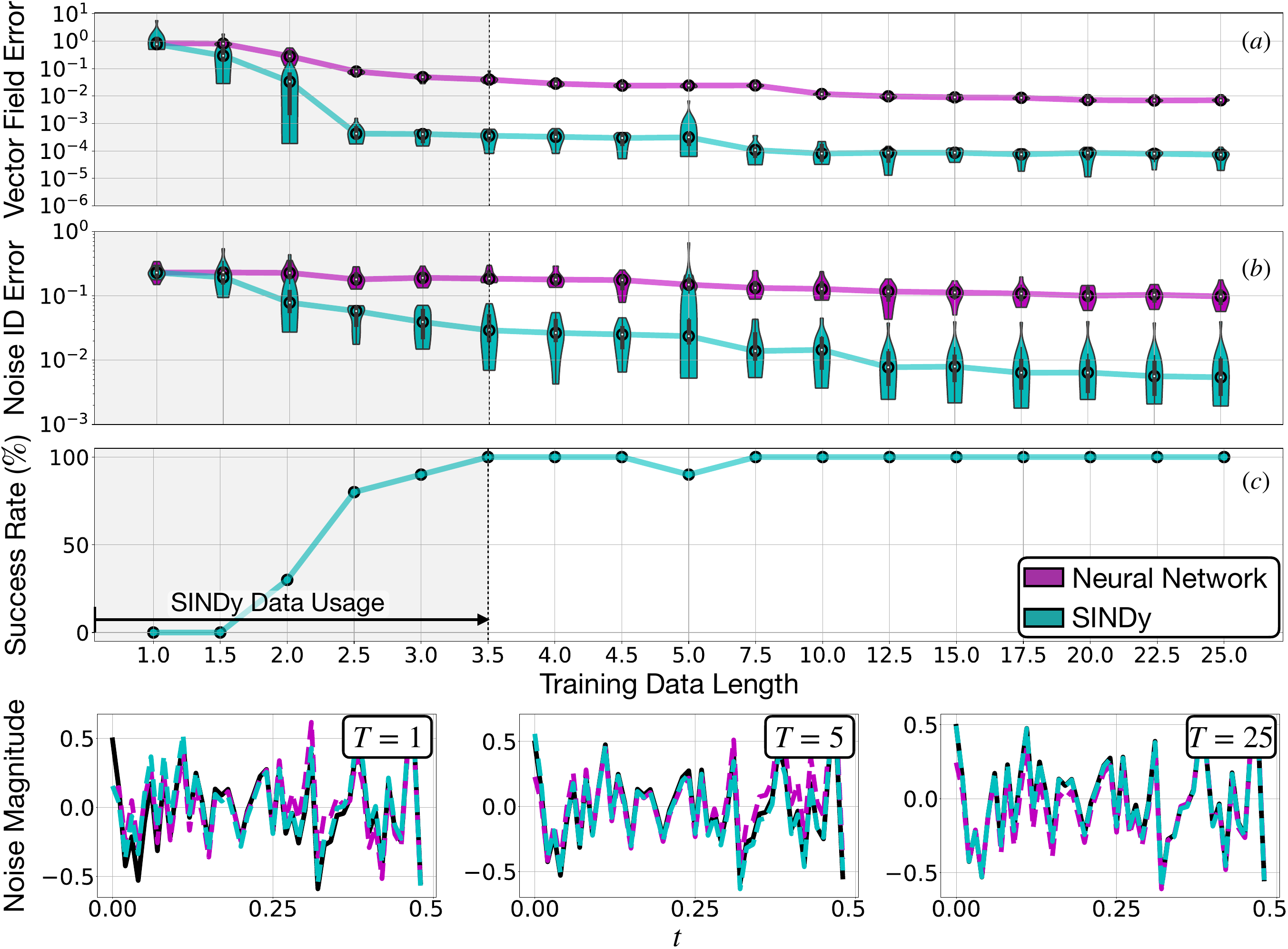}
    \caption{(a), (b): As the training data length increases, the vector field error and noise identification error of modified SINDy~(labeled as SINDy) and NN denoising approach by Rudy et al.~\cite{rudy2019deep} decreases. (c) The modified SINDy can use $3.5$ seconds of data to identify the system model with $100\%$ accuracy. There is a tiny drop in the success rate when the training data length is $5$ seconds due to the choice of a fixed thresholding parameter. By using a larger thresholding parameter, the success rate can be back to $100\%$. Bottom: A comparison of averaged true noise and identified noise by two approaches is shown.}
    \label{CompareLength}
\end{figure}
We also compare the performance of the NN denoising approach by Rudy et al.~\cite{rudy2019deep} and modified SINDy under different data usage with a fixed noise level. The minimum amount of data needed by modified SINDy to correctly identify the system model is shown by using Lorenz attractor as an example. To perform the numerical experiment, the same initial point, $x_0=[-5,5,25]$, is used to generate noise-free data of different temporal lengths. The time step is fixed at $dt=0.01$ with $10\%$ of Gaussian noise added to generate noisy training data. The success rate of modified SINDy is calculated to indicate the minimum amount of data needed to identify the correct system model. The prediction error is not shown since the simulation of the identified model in the low data limit is not stable. With a learning rate of $0.001$,  Adam is used to optimize the problem with the prediction step set to $q=3$ for both approaches. A fixed thresholding parameter $\lambda=0.1$ with $N_{loop}=6$ is used for modified SINDy and the library is constructed with up to second order terms (without constant term added).
Fig.~\ref{CompareLength} suggests that when the correct parameters and library is used for modified SINDy, it will out-perform the NN denoising approach by Rudy et al. given the same amount of data.

\section{Results}
In this section, we demonstrate the ability of modified SINDy to separate signal and noise while learning the system model.  The Van der Pol oscillator will be used as the example test case to show that modified SINDy can identify the correct distribution of the Gaussian noise added to the system. Additionally, we highlight several other examples tested with modified SINDy and summarize the performance. Furthermore, as a more advanced example, we show that modified SINDy can be used to separate non-Gaussian, non-zero mean, and non-symmetric noise distributions from the dynamics. Finally, we show how modified SINDy can be integrated to the discrepancy modeling approach~\cite{KK2019Discrepancy}.

\subsection{Van der Pol Oscillator}
\label{sec:vanderpol}
\begin{figure}[t]
    \centering
    \includegraphics[width=1\textwidth]{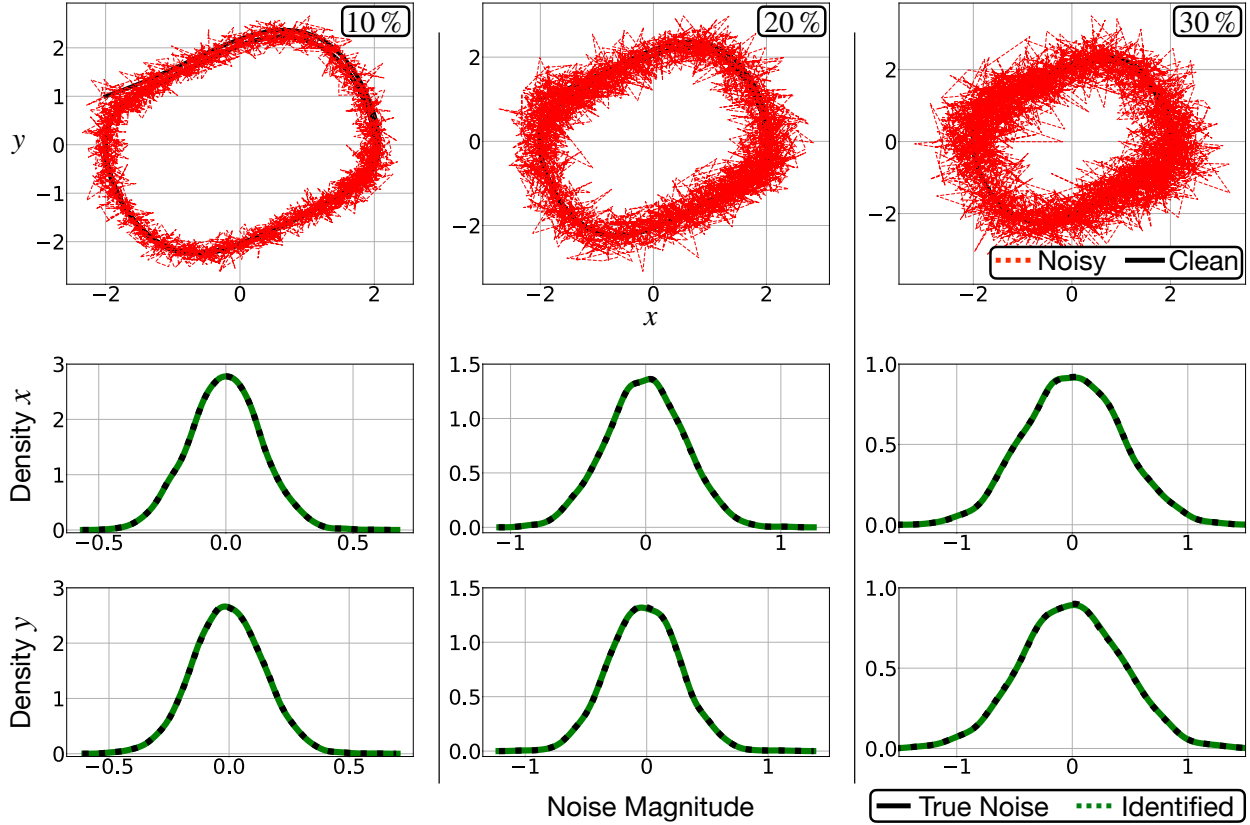}
    \caption{Distribution of the noise learned by modified SINDy is shown.}
    \label{ExampleVanderPol}
\end{figure}
The Van der Pol oscillator is used as our test case to demonstrate the ability of modified SINDy to denoise and learn the system dynamics simultaneously. The Van der Pol oscillator is given by 
\begin{equation}
    \label{eq:VanderPol}
    \begin{aligned}
        \dot{x}&=y,\\
        \dot{y}&=\mu\left(1-x^{2}\right) y-x,
    \end{aligned}
\end{equation}
where the nonlinear damping/gain parameter $\mu=0.5$ is used for demonstration purposes. The system is simulated with initial condition $[-2,1]$, $T=10$, and $dt=0.01$. The Adam optimizer with learning rate of $0.001$ is used for all noise levels. The parameters of modified SINDy are chosen as $q=1$ and $\lambda=0.05$, and the library of candidate functions is constructed with polynomial terms up to third order~(without constant term). 
Three different levels of noise are applied and the distribution of identified noise is shown in Fig.~\ref{ExampleVanderPol}. Figure~\ref{ExampleVanderPol} shows that modified SINDy correctly identified the distribution of true noise.

\subsection{Rössler Attractor}
\label{sec:rosller}
\begin{figure}[t]
    \centering
    \includegraphics[width=1\textwidth]{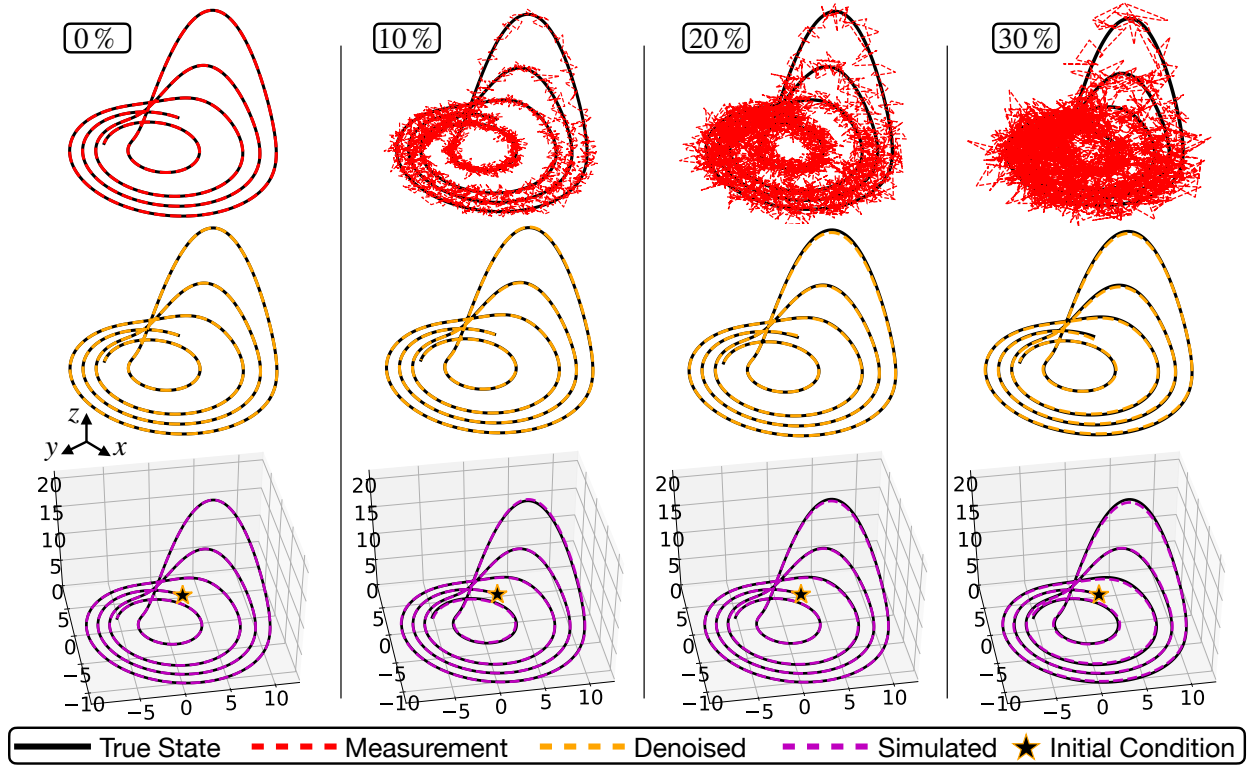}
    \caption{The modified SINDy is used to denoise the measurement of Rössler attractor while learning its model. The identified model shows high accuracy when simulating forward.}
    \label{ExampleRosller}
\end{figure}
The second example we use is the Rössler attractor that is governed by 
\begin{equation}
    \label{eq:Rossler}
    \begin{aligned}
        \dot{x}&=-y-z,\\
        \dot{y}&=x+ay,\\
        \dot{z}&=b+z(x-c),
    \end{aligned}
\end{equation}
where $a=0.2$, $b=0.2$, and $c=5.7$. The system is simulated with initial condition $[3,5,0]$, $T=25$, and $dt=0.01$. The Adam optimizer with learning rate of $0.001$ is used for all noise levels. The parameters of modified SINDy are chosen as $q=1$ and $\lambda=0.05$, and the library of candidate functions is constructed with polynomial terms up to second order~(with constant term). Three different levels of noise are applied and the denoised signal is shown in Fig.~\ref{ExampleRosller}. Figure~\ref{ExampleRosller} also shows the simulated trajectories of the identified models. The initial condition $[3,5,0]$, $T=25$, and $dt=0.01$ are used to simulate the identified models.

\subsection{Lorenz 96 Model}
\label{sec:lorenz96}
\begin{figure}[t]
    \centering
    \includegraphics[width=1\textwidth]{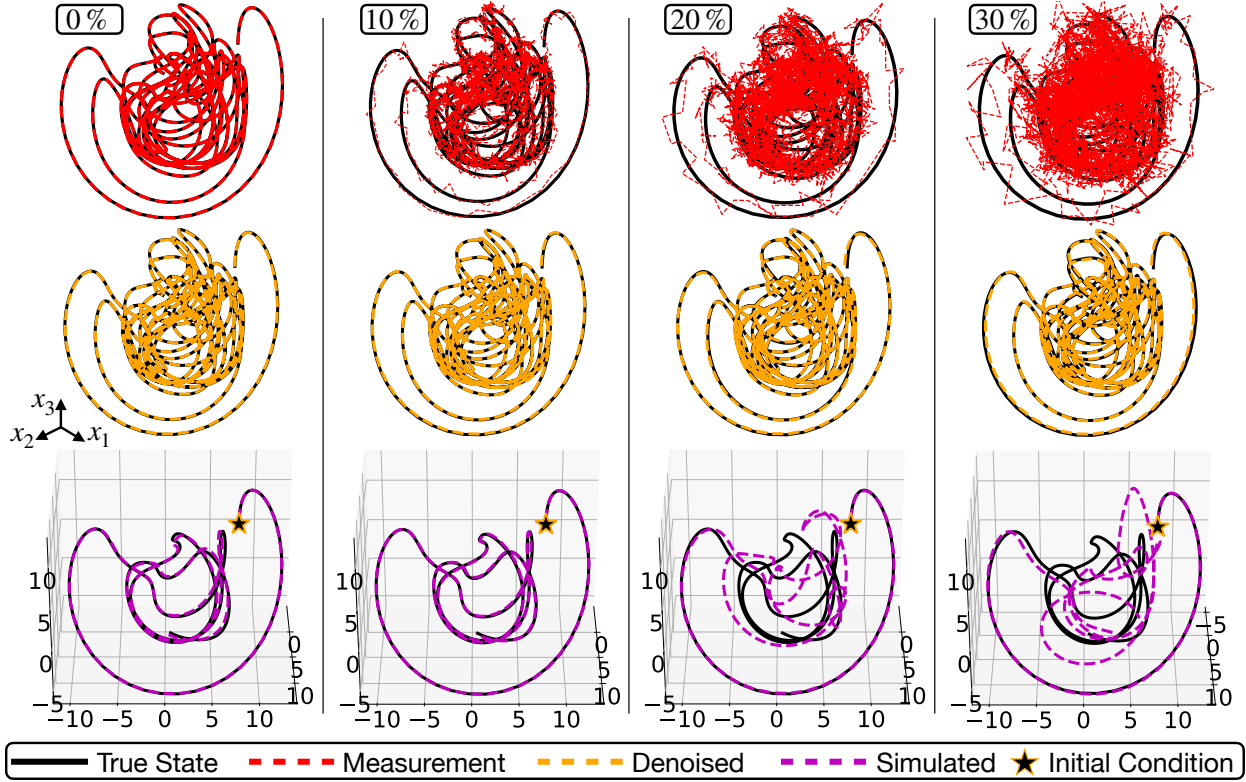}
    \caption{The modified SINDy is used to denoise the measurement of Lorenz 96 system while learning its model.}
    \label{ExampleLorenz96}
\end{figure}
As our last example, we use the modified SINDy to identify Lorenz 96 model whose equation is given by 
\begin{equation}
    \label{eq:Loren96}
        \dot{x}_i=(x_{i+1}-x_{i-2})x_{i-1}-x_{i}+F,
\end{equation}
for $i=1, 2, \dots, N$. We assume $x_{-1}=x_{N-1}$, $x_0=x_{N}$, $x_{1}=x_{N+1}$, and set forcing term $F$ as $8$ to generate chaotic behavior. The number $N$ is set as $4$ such that the model has $6$ states. The system is simulated with initial condition $[1,8,8,8,8,8]$, $T=25$, and $dt=0.01$. The Adam optimizer with learning rate of $0.001$ is used for all noise levels. The parameters of modified SINDy are chosen as $q=1$ and $\lambda=0.1$~(for $30\%$ noise, $\lambda=0.05$). The library of candidate functions is constructed with polynomial terms up to third order~(with constant term included, $84$ candidates in total). Three different levels of noise are applied and the denoised signal is shown in Fig.~\ref{ExampleLorenz96}~(for ease of visualization, only the first three states are shown). Figure~\ref{ExampleLorenz96} also shows the simulated trajectories of identified models. The initial condition $[1,8,8,8,8,8]$, $T=5$, and $dt=0.01$ are used to simulate the identified models.

\begin{figure}[t]
    \centering
    \includegraphics[width=1\textwidth]{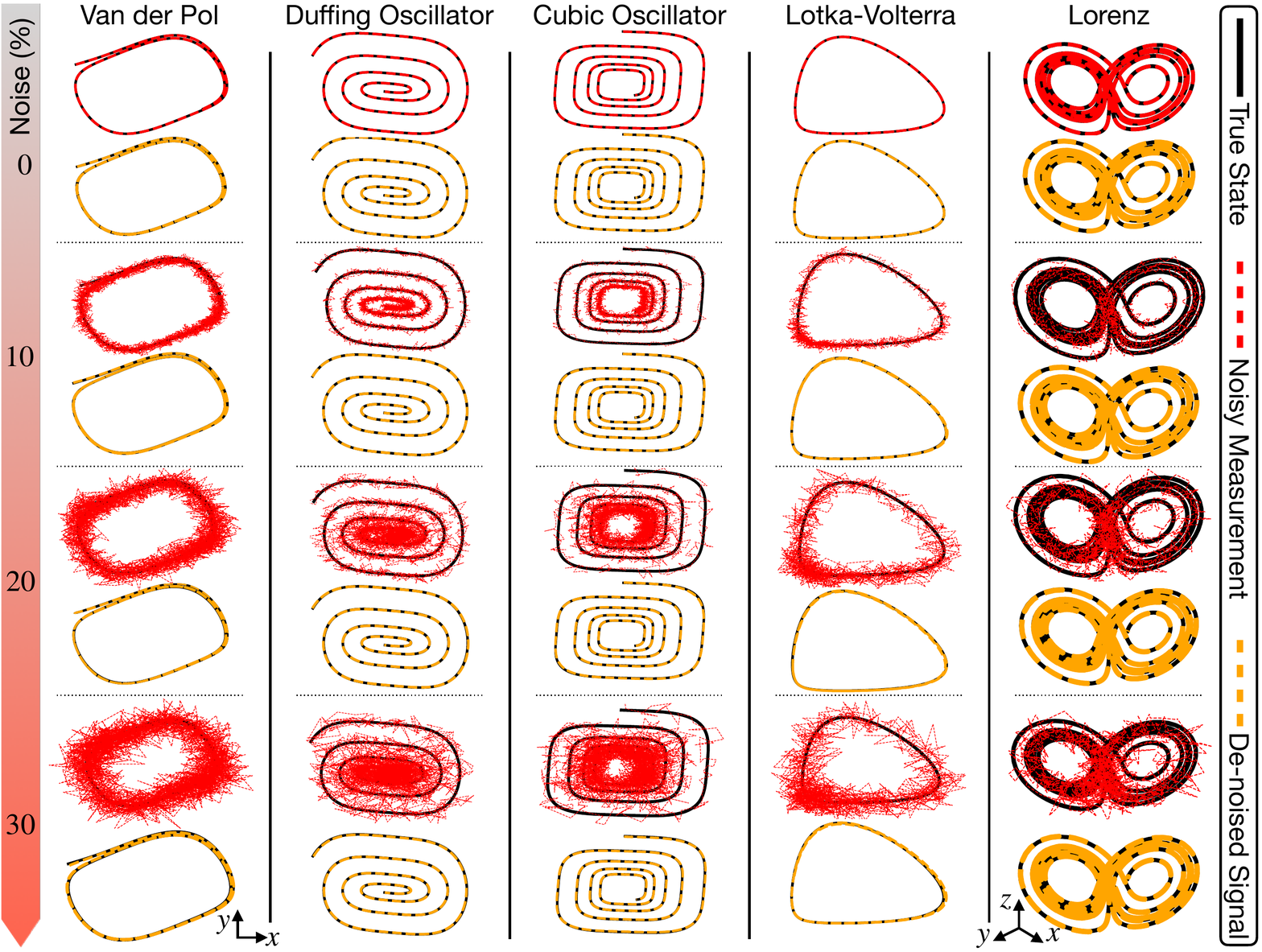}
    \caption{This plot shows the denoising ability of modified SINDy with different examples. Regardless of the noise percentage, modified SINDy correctly identified all the system models.}
    \label{SumExamples}
\end{figure}
In Fig.~\ref{SumExamples}, the effectiveness of modified SINDy is demonstrated on a number of canonical dynamical systems models. For all examples,  Gaussian noise with zero-mean is added to generate the noisy training data, and Adam optimizer is used to perform the optimization. The models and other parameters used for each example are summarized in Appendix.~\ref{Appendix:ExampleDetails}. The modified SINDy correctly identified all the system model and noise distribution regardless of the noise magnitude used.

\subsection{Identification of Noise Distributions}
\label{sec:DiffNoise}
\begin{figure}[t]
    \centering
    \includegraphics[width=1\textwidth]{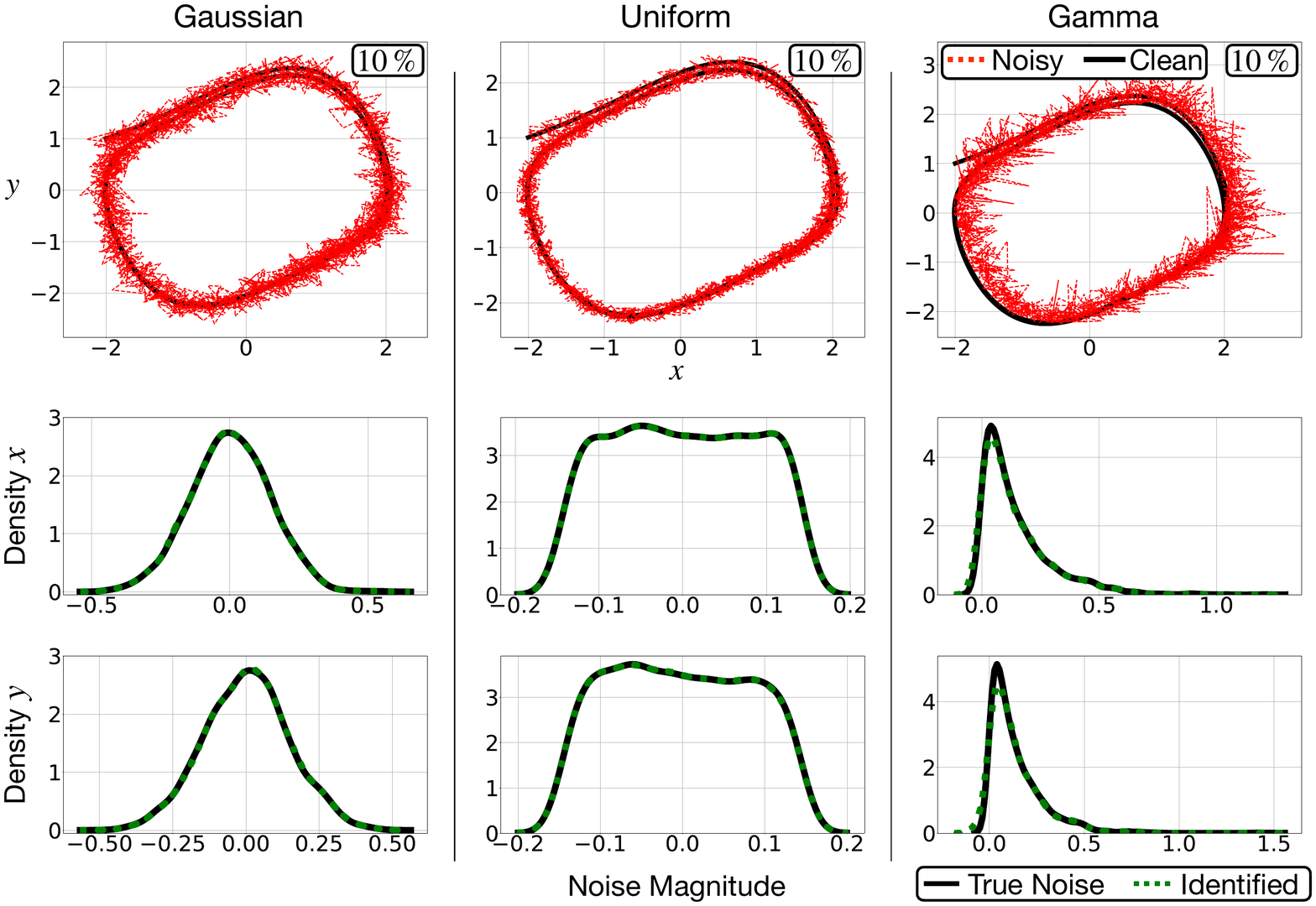}
    \caption{The ability of modified SINDy to handle different kinds of noise distribution is illustrated in this figure, and the Van der Pol oscillator is used as an example. $10\%$ of noise is generated and added to the clean signal. As this figure shows, modified SINDy can identify different types of noise distribution correctly.}
    \label{ThreeDisNoise}
\end{figure}

The modified SINDy algorithm has the ability to to handle different kinds of noise distributions. Three different kinds of noise distributions are used to demonstrate this: Gaussian, Uniform, and Gamma. To generate the Gamma noise, its shape and scale are set to $1$. The generated noise is multiplied by $\text{Noise Percentage}\times \mbox{var}(\text{Signal})$. The noise-free data of Van der Pol oscillator is generated the same way in Sec.~\ref{sec:vanderpol}. The prediction step is set to $q=2$ and the sparsity parameter is set to $\lambda=0.15$. Figure~\ref{ThreeDisNoise} shows the distribution identified by modified SINDy. Figure~\ref{ThreeDisNoise} shows that learning the non-zero mean noise distribution is more difficult than learning a zero-mean one. For better learning results of a non-zero mean noise distribution, one can try the iterative learning approach shown in Appendix.~\ref{Appendix:NonZeroNoise}. Once the noise is separated from the signal, an additional step can be taken to identify the distribution of noise from the candidate distributions. This can be achieved by the \texttt{fitter} package in Python~\cite{Cokelaer2019fitter}. Appendix.~\ref{Appendix:NoiseType} shows more details of this process.

\subsection{Discrepancy Modeling}
\begin{figure}[t]
    \centering
    \includegraphics[width=1\textwidth]{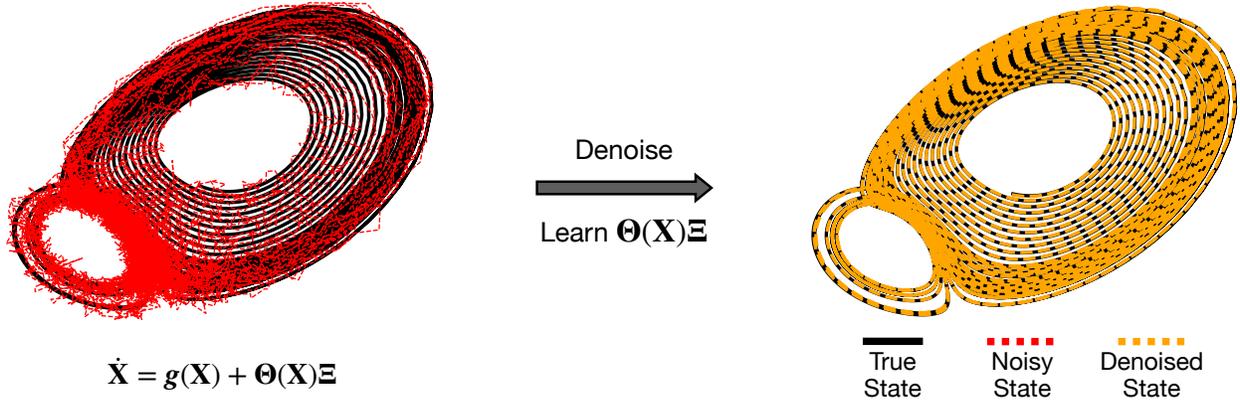}
    \caption{The modified SINDy can be easily integrated with the discrepancy learning method. The known structure $\bsg(\bX)$ can be utilized while we aim to learn the discrepancy and denoising the measurement.}
    \label{FigDiscrepancy}
\end{figure}

Modified SINDy can be easily integrated with the discrepancy modeling framework of SINDy ~\cite{KK2019Discrepancy}. This is of great practical value since it is often the case that parts of the dynamics is known. Suppose the known~(theoretical) right-hand side dynamics in Eq.~(\ref{eq:dx=f(x)}) is $\bsg(\bx)$.  Discrepancy modeling assumes that the known model is not capable of modeling the data due to missing physics terms on the right-hand side.  Thus there
is a mismatch between the derivative $\dot{\bx}$ and the known dynamics $\bsg(\bx)$.  The  discrepancy modeling approach tries to identify the missing dynamics $\bTheta(\bx)\bXi$ such that
\begin{equation}
    \label{eq:discrepancy}
    \dot{\bx}=\bsf(\bx)=\bsg(\bx)+\bTheta(\bx)\bXi.
\end{equation}
To illustrate this process, consider a system $\dot{\bx}=\bsf(\bx)$, whose model is given as 
\begin{equation}
    \label{eq:ModifiedLorenz}
    \begin{aligned}
        \dot{x}&= -10x+10y+xy,\\
        \dot{y}&= 28x-xz-y+3z,\\
        \dot{z}&= xy-8/3z.
    \end{aligned}
\end{equation}
Eq.~\eqref{eq:ModifiedLorenz} is simulated with the $x_0=[5,5,25]$, $T=30$, and $dt=0.005$ to generate noise-free data. Training data is produced by adding $10\%$ Gaussian noise in order to create the noisy measurement. Assume that the noisy measurement of Eq.~\eqref{eq:ModifiedLorenz} is given.  Further assume that the dynamics is modified based on $\bsg(\bx)$, which is given by
\begin{equation}
    \label{eq:Lorenz}
    \begin{aligned}
        \dot{x}&= -9.5x+10.5y,\\
        \dot{y}&= 27.6x-1.1xz-0.9y,\\
        \dot{z}&= 1.05xy-2.6z.
    \end{aligned}
\end{equation}
The difference between the Eq.~\eqref{eq:ModifiedLorenz} and Eq.~\eqref{eq:Lorenz} will be the discrepancy model $\bTheta(\bx)\bXi$ that modified SINDy identifies. Note that this prior information of the dynamics, $\bsg(\bx)$, can be constrained to exist in the modified SINDy library during the optimization process, and its parameters can be used as an initial guess of the true parameters. Thus, the only thing we have to learn is the missing dynamics. Figure~\ref{FigDiscrepancy} illustrates this process. In this example, the $q=4$, $\lambda=0.4$, and the learning rate of Adam optimizer is $0.001$. Fig.\ref{FigDiscrepancy} suggests that modified SINDy can be used to learn the discrepancy model when parts of the dynamics are already known.

\section{Conclusion and Future Work}
In this work, we introduce a new learning algorithm that leverages automatic differentiation and sparse regression for simultaneously (i) denoising time-series data, (ii) learning and parametrizing the noise probability distribution, and (iii) identifying the underlying parsimonious dynamical system responsible for generating the time-series data.  The method provides a critically enabling modification to the SINDy algorithm for improving robustness to noise with less training data in comparison with the previously developed NN denoising approach by Rudy et al.~\cite{rudy2019deep}. Multiple numerical examples are shown to demonstrate the effectiveness of the modified SINDy method for signal and noise separation as well as model identification.  It is also shown that the modified SINDy can be integrated with a discrepancy modeling framework whereby prior information of the dynamical model can be used to help identify the missing dynamics. Importantly, we have shown that modified SINDy can be used to learn various types of noise distributions, including Gaussian, uniform, and non-zero mean noise distributions, such as a Gamma distribution.  Overall, the modified SINDy is a robust method with practical potential for handling highly noisy data sets and/or when partial model information is known. 

The modified SINDy is modular, allowing for many easily integrated improvements.  An important direction for development includes the incorporation of control inputs, since many systems of practical interest are actuated, such as the pendulum on a cart system~\cite{Brunton2016SINDyc,kaheman2020sindypi}. Extending modified SINDy to consider the impact of control will significantly expand its application domain. Improvements in computational speed are also desirable.  In comparison with the sequential least-square thresholding of the standard SINDy algorithm, the Adam optimizer is slow. There is the potential to use the standard SINDy sparse regression algorithms to warm start the Adam optimization routine. The modified SINDy can also be integrated with SINDy-PI to identify rational or implicit dynamics, which is quite difficult since the simulation error shown in Eq.~\eqref{eq:minSimAllPoint} can not be calculated easily when the dynamics take a rational form. This is the case where the use of the NN denoising approach~\cite{rudy2019deep} by Rudy et al. is ideal.

Finally, it is important to improve the robustness of the modified SINDy algorithm when a large number of library terms are used. Currently, the modified SINDy can not handle large libraries robustly due to the non-convexity of the optimization problem. When the library is too large, the problem becomes unstable without decreasing the optimizer's learning rate. One potential solution is to simulate the dynamics with a variable time step numerical simulation scheme instead of a fixed step scheme, as we used in this paper. Although there are still many improvements to be made, we believe the introduction of modified SINDy will help guide the use of automatic differentiation tools to improve the SINDy framework.

\section*{Acknowledgments}
SLB acknowledges support from the Army Research Office (ARO W911NF-19-1-0045) and the Air Force Office of Scientific Research (AFOSR FA9550-18-1-0200). 
JNK acknowledges support from the Air Force Office of Scientific Research (AFOSR FA9550-17-1-0329). 
We also acknowledge valuable discussions with Samuel H. Rudy, Jared Callaham, Henning Lange, Daniel A. Messenger, and Benjamin Herrmann.

\newpage

\begin{spacing}{.8}
    \small{
    \setlength{\bibsep}{1.5pt}
    \bibliographystyle{IEEEtran}
    \bibliography{PaperReference}
    }
\end{spacing}

\newpage
\appendix
\section{Algorithm for Simultaneously Denoising and Learning System Model}
\label{Appendix:Algorithm}
\begin{algorithm}[H]
\SetKwInput{KwInput}{Input}                
\SetKwInput{KwOutput}{Output} 

\SetCommentSty{mycommfont}

\DontPrintSemicolon
\KwInput{$\bY,\ \bTheta(*),\ dt,\ \lambda,\ N_{loop},\ \omega$}
\KwOutput{$\bXi$, $\bhN$}
\tcc{Initialize the value of $\bhN$}
\eIf{$\mathrm{Soft Start}$}{
  $\bhN=\bY-\mathrm{smoothSignal}(\bY)$\tcp*{If the soft start is true, the estimated value of noise is obtained by pre-smoothing the noisy signal.}
  }{
  $\bhN=\mathrm{zeros}(\mathrm{size}(\bY))$\tcp*{Else, the estimated value of noise is initialized using zero matrix.}
  }
  
\tcc{Initialize the value of $\bXi$}
$\bhX=\bY-\bhN$.

Calculate $\bdhX$ using $\bhX$.

$\bXi=\text{SINDy}(\bdhX,\bTheta(\bhX),\lambda)$.

\tcc{Simultaneously denoising and learning system model}
\While{$k<N_{loop}$}{
    Optimize $\mathcal{L}(\bXi,\bhN)$ shown in Eq.~\eqref{eq:optProb}.
    
    $(|\bXi|<\lambda) =\boldsymbol{0}$.\tcp*{Constrain the elements in $\bXi$ whose absolute value smaller than $\lambda$ as zero during the rest of optimization.}
    
    $\bhX=\bY-\bhN$.\tcp*{Get new estimate of true state.}
    
    Calculate $\bdhX$ using $\bhX$.\tcp*{Get new estimate of true derivative.}
    
    $(|\bXi|\neq \boldsymbol{0})=\bTheta(\bhX)\backslash\bdhX$.\tcp*{Regress the dynamics on terms in $\bXi$ that are not constrained as zero.}
 }
\caption{Modified SINDy}
\end{algorithm}

\section{Effect of Thresholding Parameter $\lambda$}
\label{Appendix:EffectOfLambda}
\begin{figure}[t]
    \centering
    \includegraphics[width=\textwidth]{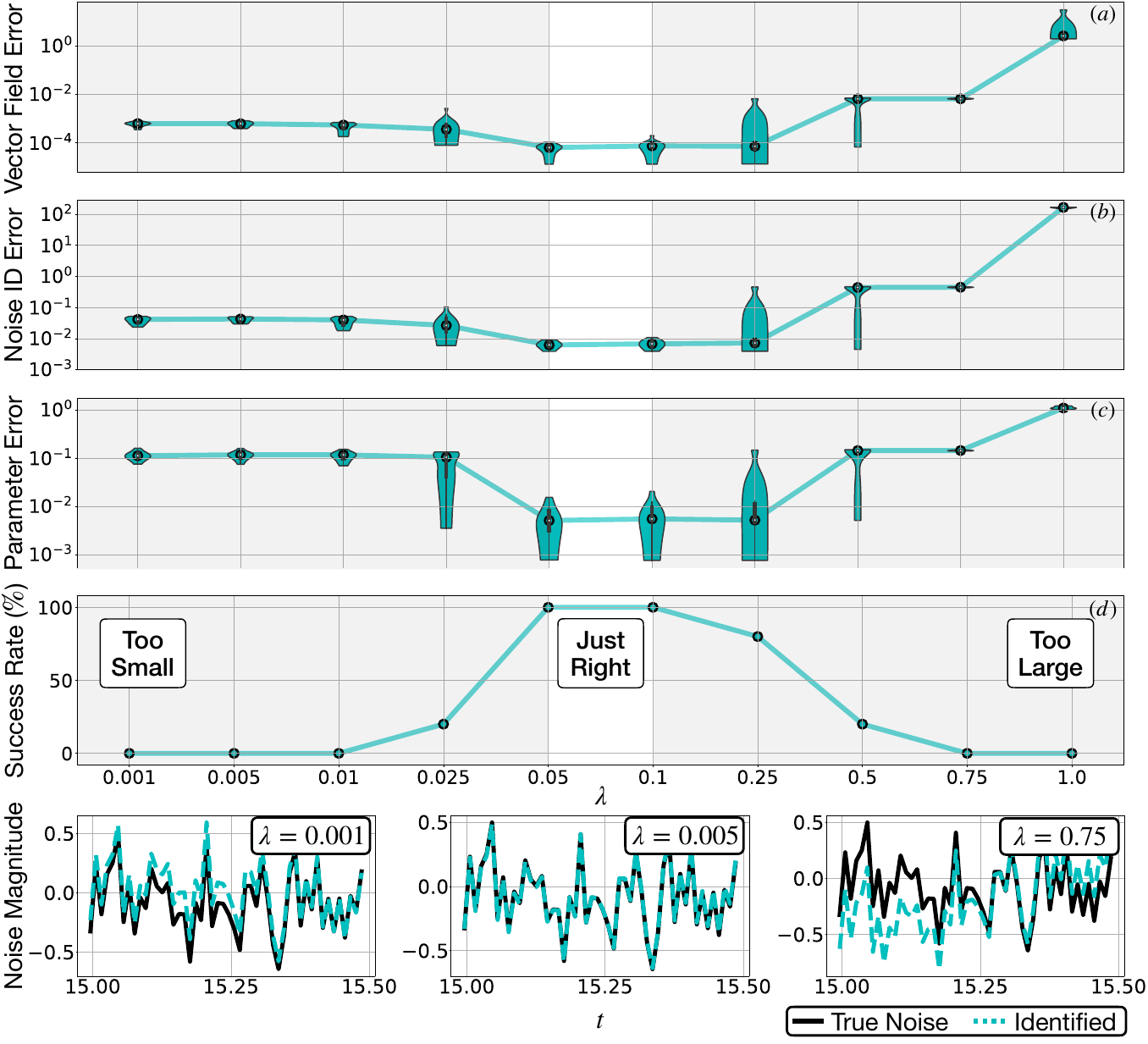}
    \caption{This figure shows how the choice of sparsity parameter $\lambda$ will effect modified SINDy performance.  If $\lambda$ is too small, modified SINDy will not converge to the correct model in a short range of time and will be stuck in the local minimum. On the contrary, if the sparsity parameter is too large, the identified model will miss the necessary term to build the correct model. Thus, the value of $\lambda$ needs to be tuned properly to determine the accurate model.}
    \label{EffectLambda}
\end{figure}

Thresholding parameter $\lambda$ is the most important parameter to tune in modified SINDy. The parameter $\lambda$ will determine the sparsity of the model structure. It's effect can be seen in Fig.~\ref{EffectLambda}. In Fig.~\ref{EffectLambda}, Lorenz equation is simulated with $[-5.0,5.0,25.0]$, $dt=0.01$, and $T=25$. $10\%$ of Gaussian noise is added and Adam optimizer with learning rate of $0.001$ is used to denoise the signal. $N_{loop}$ is set to $8$ and different values of $\lambda$ is used. For each $\lambda$, the numerical experiments is performed $10$ times to calculate the median and distribution of the error as shown in Fig.~\ref{EffectLambda}. Fig.~\ref{EffectLambda} suggests that the value of $\lambda$ must be properly tuned. If the value of $\lambda$ is too small, the sparsity constraint will not be strong enough to enforce the correct model to be found. Moreover, $\bXi$ and $\bhN$ will easily get stuck in the local minimum. If the value of $\lambda$ is too large, the correct terms can be wrongly eliminated and the resulting model structure will be wrong. If the model structure is wrong, there will be huge difference between the identified noise $\bhN$ and true noise $\bN$. To avoid swiping different values of $\lambda$, our proposed method can be easily modified to use the stepwise sparse regression~(SSR) approach~\cite{boninsegna2018sparse}. However, the use of SSR approach and its performance is not in the scope of this paper.

\section{Effect of Prediction Step $q$}
\label{Appendix:EffectOfPreStep}
\begin{figure}[t]
    \centering
    \includegraphics[width=1\textwidth]{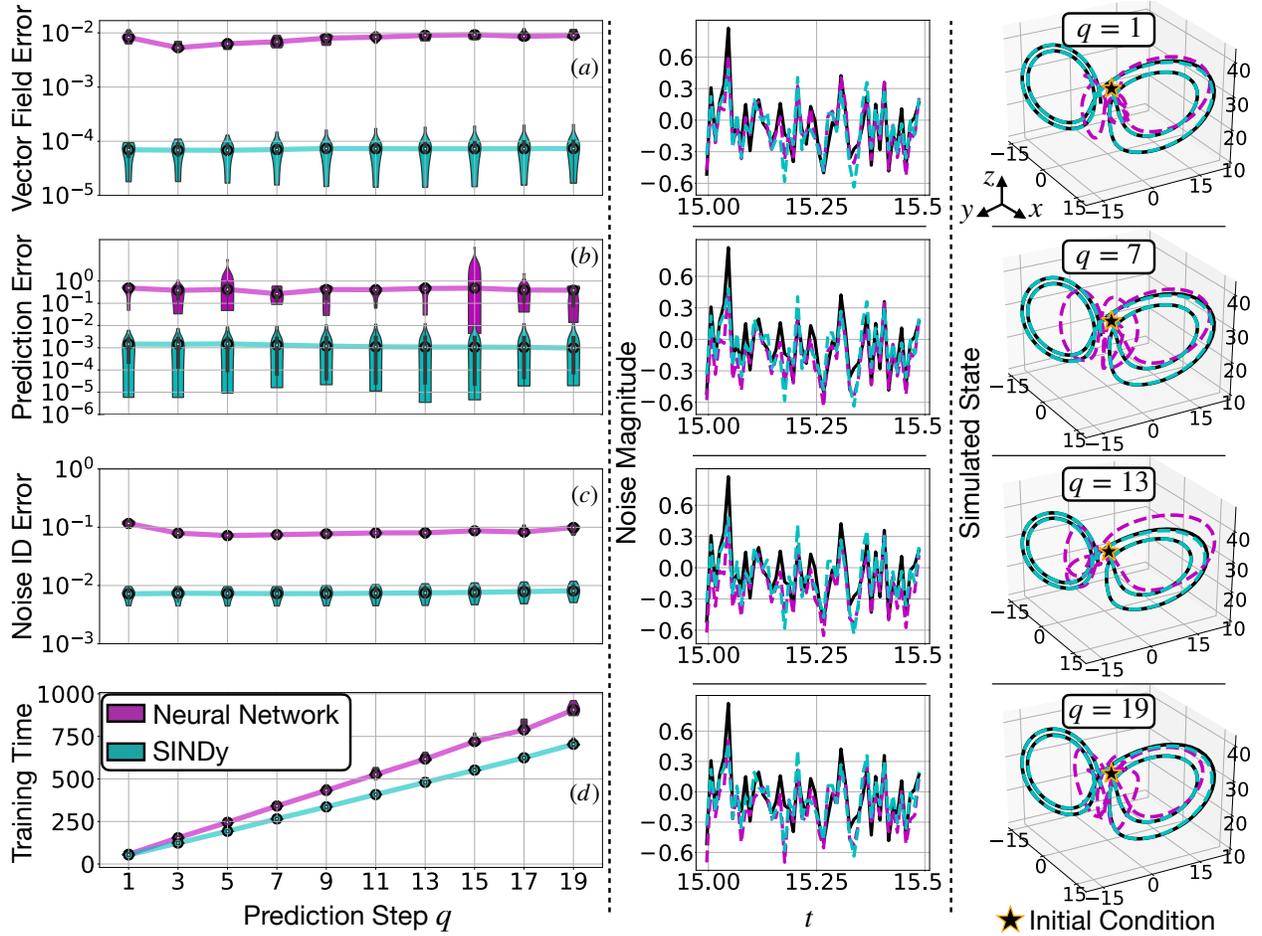}
    \caption{Left: The value of noise identification error, prediction error, and vector field error of NN denoising approach by Rudy et al.~\cite{rudy2019deep} and modified SINDy~(labeled as SINDy) are shown as the value of $q$ changes. In (a) to (d), the black circle shows the median of the error of $10$ runs while the violin shape represents the distribution of calculated result. Mid: The average of true noise and identified noise of modified SINDy and NN approach is shown for four different prediction steps. Right: The comparison of the simulated and true trajectory of modified SINDy and NN identified model is shown. The simulated trajectory uses $x_0=[-5,5,25]$ and $dt=0.01$. The model is simulated for $3$ seconds. It could be seen that modified SINDy has better performance in this case. All the computation is performed on RTX 2080 GPU, with 32GBs of RAM and AMD Ryzen 7 2700X Processor.}
    \label{CompareQ}
\end{figure}

Fig.~\ref{CompareQ} shows the effect of the prediction step $q$ on the performance of NN denoising approach by Rudy et al.~\cite{rudy2019deep} and modified SINDy approach. The chaotic Lorenz system is used for comparison. The Lorenz attractor is simulated by setting $x_0=[-5,5,25]$, $T=25$, and $dt=0.01$. The noise level is set to $10\%$ to generate noisy data. Each prediction step is run for $10$ times to calculate the median of the error. Adam optimizer, with a learning rate of $0.001$ is used to perform the optimization. $N_{loop}$ is set to $3$. Fig.~\ref{CompareQ} suggests that the performance of modified SINDy is not hugely affected by the prediction step $q$. However, for the NN denoising approach shown in~\cite{rudy2019deep}, there exist some value of $q$ to achieve optimal performance. Fig.~\ref{CompareQ} also suggests the computational time of both approaches increase linearly as the value of $q$ increase. Thus, $q$ can be chosen as a small value to save the computational time when using modified SINDy without sacrificing too much of the performance.

\section{Effect of Optimization Iteration $N_{loop}$}
\label{Appendix:EffectOfNloop}
\begin{figure}[t]
    \centering
    \includegraphics[width=1\textwidth]{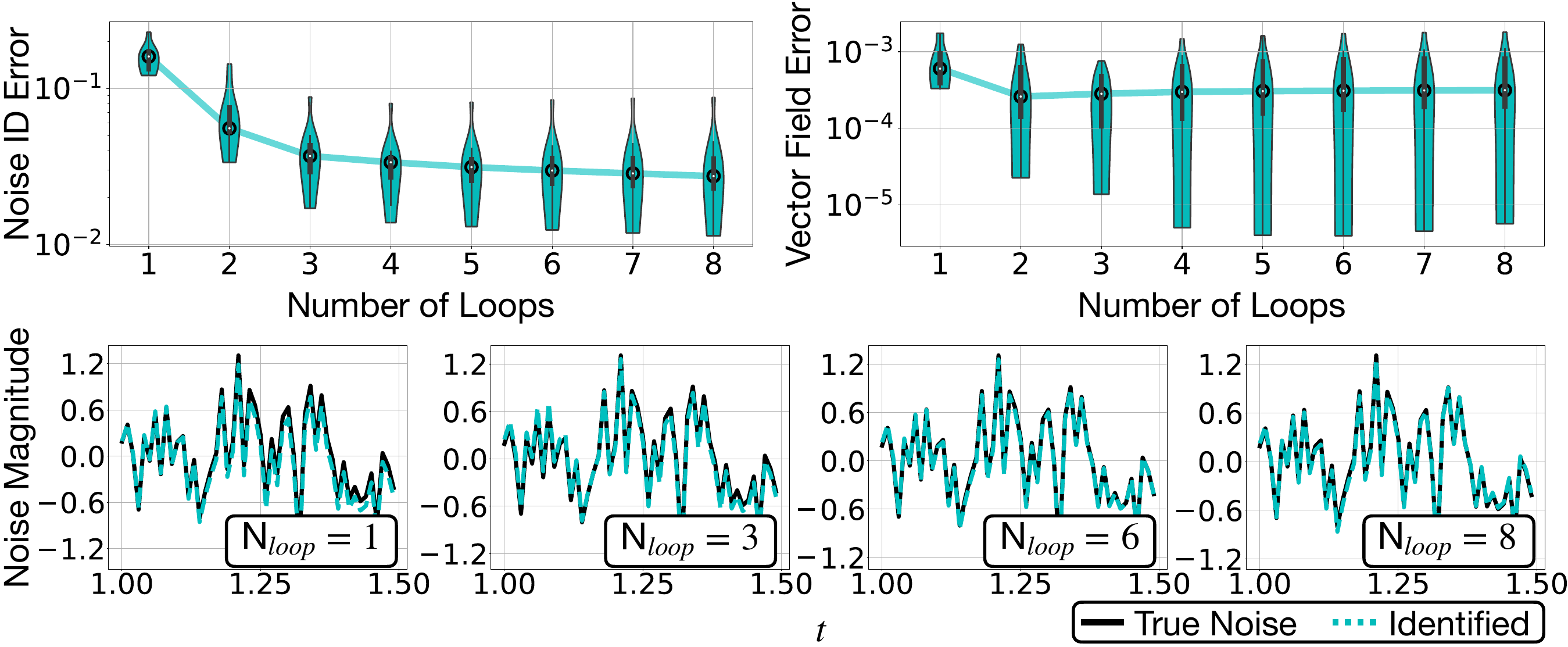}
    \caption{Top Left: As the number of optimization loop increases, the noise identification error asymptotically decreases and converges to the optimal value. The black circle indicates the median of $10$ runs while the violin shape represents error distribution. Top Right: As the number of loop increases, the vector field error gradually converges to a certain value. Bottom: The average of the identified noise and true noise is shown for four different choice of optimization loops. As the number of loop increases, the differences between the true noise and identified noise is minimized.}
    \label{CompareNloop}
\end{figure}

The parameter $N_{loop}$ determines how many times the thresholding optimization is performed. Fig.~\ref{CompareNloop} shows the effect of $N_{loop}$ on the noise identification error and vector field error using Lorenz attractor as an example. The system is simulated by setting $x_0=[5,5,25]$, $T=25$, $dt=0.01$, and $q=3$. Adam optimizer, with a learning rate of $0.001$, is used to optimize the problem. Fig.~\ref{CompareNloop} suggests the performance of modified SINDy will gradually converge in the end.

\begin{figure}[t]
    \centering
    \includegraphics[width=1\textwidth]{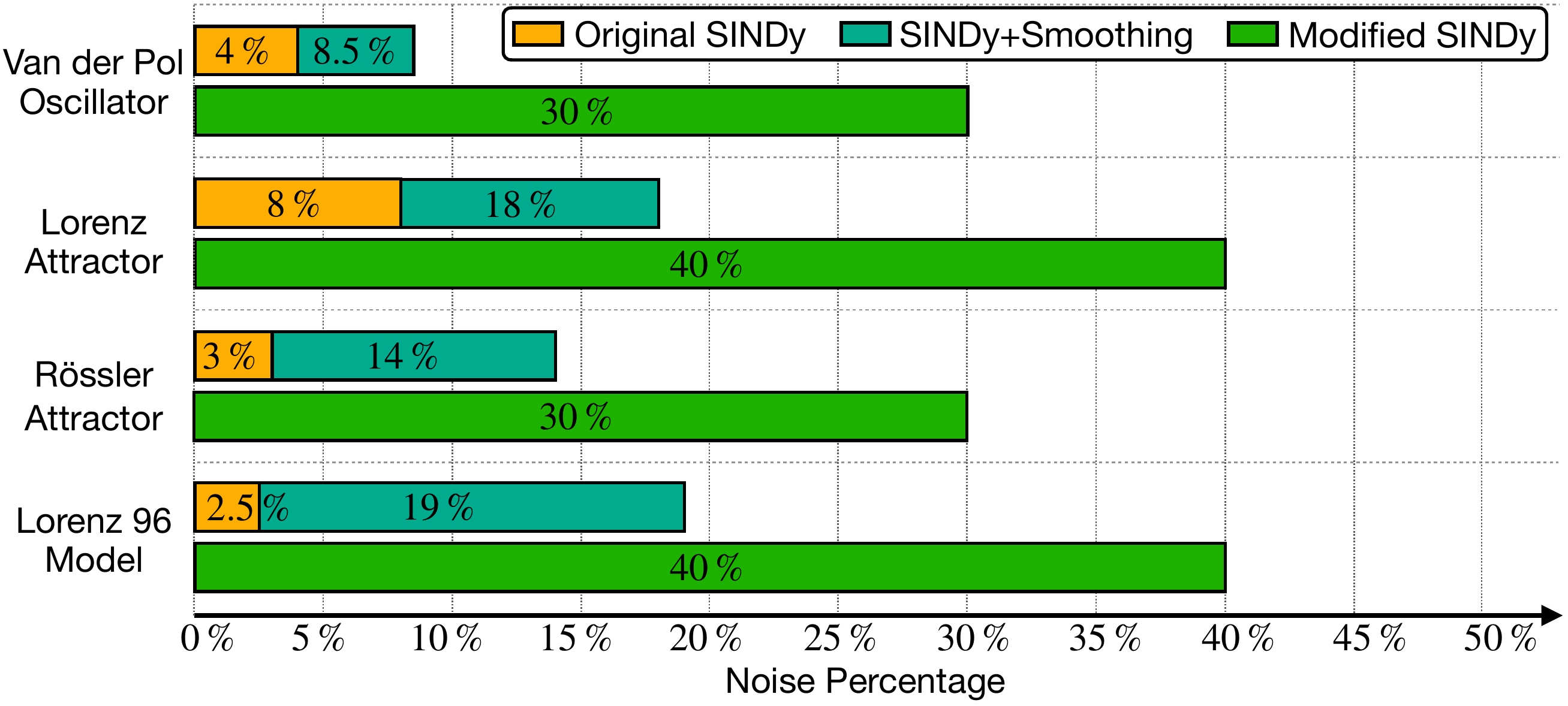}
    \caption{The maximum level of noise SINDy and modified SINDy can handle to generate the correct model structure is shown. Tikhonov regularization approach is used to pre-smooth the noisy data. It can be seen that the modified SINDy is about $2$ times more robust than original SINDy~\cite{Brunton2016SINDy}.}
    \label{CompareSINDy}
\end{figure}
\section{Noise Robustness Comparison with SINDy}
\label{Appendix:SINDy}
This section shows the noise robustness comparison of SINDy~\cite{Brunton2016SINDy} and modified SINDy using Van der Pol oscillator, Lorenz attractor, and Rössler attractor. Fig.~\ref{CompareSINDy} shows the maximum noise percentage each algorithm can handle to generate the correct model structure. For each noise level, $5$ different noisy data sets are generated and provided to both approaches. If the tested algorithm fails to identify the correct model structure for any noisy data sets at a given noise level, we will assume it is not robust to noise at this level. For SINDy, the derivative is computed using finite difference, and we show the effect of pre-smoothing the noisy data on its performance. Note that no smoothing is applied for modified SINDy. The clean data for Lorenz attractor, Van der Pol oscillator, and Rössler attractor is generated the same way shown in Sec.~\ref{sec:CompareNoiseNNandModifiedSINDy}, Sec.~\ref{sec:vanderpol}, and Sec.~\ref{sec:rosller}. For SINDy, the sparsity parameter $\lambda$ is chosen as a hundred uniformly distributed values from $0.01$ to the minimum of true parameters' absolute value. For modified SINDy, $q=1$ and $N_{loop}=8$ are used for all examples shown in Fig.~\ref{CompareSINDy}. Table.~\ref{Table:CompareSINDy} shows other parameters we used for modified SINDy. Note that it is possible to make modified SINDy work at a higher noise level by tuning its parameters. However, swiping various parameters is quite computationally heavy for modified SINDy. Thus, the maximum noise level modified SINDy can tolerate in Fig.~\ref{CompareSINDy} is a lower approximate.
\begin{table}[]
\centering
\caption{Parameters used for modified SINDy in Fig.~\ref{CompareSINDy} under maximum noise it can tolerate. The constant term is included when building the library for Rössler attractor and Lorenz 96 model but not for other examples. The parameter error is calculated using Eq.~\eqref{eq:Ep}.}
\label{Table:CompareSINDy}
\begin{tabular}{|c||c|c|c|c|c|c|c|c|}
\hline
Model                        & \begin{tabular}[c]{@{}c@{}}Noise \\ Percentage\end{tabular} & \begin{tabular}[c]{@{}c@{}}Library\\  Order\end{tabular} & \begin{tabular}[c]{@{}c@{}}Random \\ Seed\end{tabular}        & $0$      & $1$     & $2$     & $3$      & $4$      \\ \hline\hline
\multirow{3}{*}{Lorenz}      & \multirow{3}{*}{$30\%$}                                     & \multirow{3}{*}{$2$}                                     & $\lambda$                                                     & $0.3$    & $0.2$   & $0.3$   & $0.1$    & $0.1$    \\ \cline{4-9} 
                             &                                                             &                                                          & \begin{tabular}[c]{@{}c@{}}Parameter \\ Error\end{tabular}    & $0.0046$ & $0.051$ & $0.031$ & $0.032$  & $0.077$  \\ \cline{4-9} 
                             &                                                             &                                                          & \begin{tabular}[c]{@{}c@{}}Max Adam \\ Iteration\end{tabular} & $10000$  & $15000$ & $15000$ & $15000$  & $15000$  \\ \hline\hline
\multirow{3}{*}{Rössler}     & \multirow{3}{*}{$40\%$}                                     & \multirow{3}{*}{$2$}                                     & $\lambda$                                                     & $0.1$    & $0.22$  & $0.22$  & $0.1$    & $0.1$    \\ \cline{4-9} 
                             &                                                             &                                                          & \begin{tabular}[c]{@{}c@{}}Parameter \\ Error\end{tabular}    & $0.021$  & $0.059$ & $0.022$ & $0.0062$ & $0.020$  \\ \cline{4-9} 
                             &                                                             &                                                          & \begin{tabular}[c]{@{}c@{}}Max Adam \\ Iteration\end{tabular} & $15000$  & $15000$ & $15000$ & $15000$  & $15000$  \\ \hline\hline
\multirow{3}{*}{Van der Pol} & \multirow{3}{*}{$30\%$}                                     & \multirow{3}{*}{$3$}                                     & $\lambda$                                                     & $0.1$    & $0.22$  & $0.22$  & $0.1$    & $0.1$    \\ \cline{4-9} 
                             &                                                             &                                                          & \begin{tabular}[c]{@{}c@{}}Parameter \\ Error\end{tabular}    & $0.053$  & $0.039$ & $0.014$ & $0.011$  & $0.041$  \\ \cline{4-9} 
                             &                                                             &                                                          & \begin{tabular}[c]{@{}c@{}}Max Adam \\ Iteration\end{tabular} & $15000$  & $15000$ & $15000$ & $15000$  & $5000$   \\ \hline\hline
\multirow{3}{*}{Lorenz $96$} & \multirow{3}{*}{$40\%$}                                     & \multirow{3}{*}{$3$}                                     & $\lambda$                                                     & $0.1$    & $0.15$  & $0.09$   & $0.215$  & $0.1$    \\ \cline{4-9} 
                             &                                                             &                                                          & \begin{tabular}[c]{@{}c@{}}Parameter\\ Error\end{tabular}     & $0.015$  & $0.015$ & $0.016$ & $0.034$  & $0.0075$ \\ \cline{4-9} 
                             &                                                             &                                                          & \begin{tabular}[c]{@{}c@{}}Max Adam \\ Iteration\end{tabular}  & $7000$   & $7000$  & $10000$  & $10000$  & $5000$   \\ \hline
\end{tabular}
\end{table}

\section{Noise Robustness Comparison with Weak-SINDy}
\label{Appendix:WeakSINDy}
\begin{figure}[t]
    \centering
    \includegraphics[width=1\textwidth]{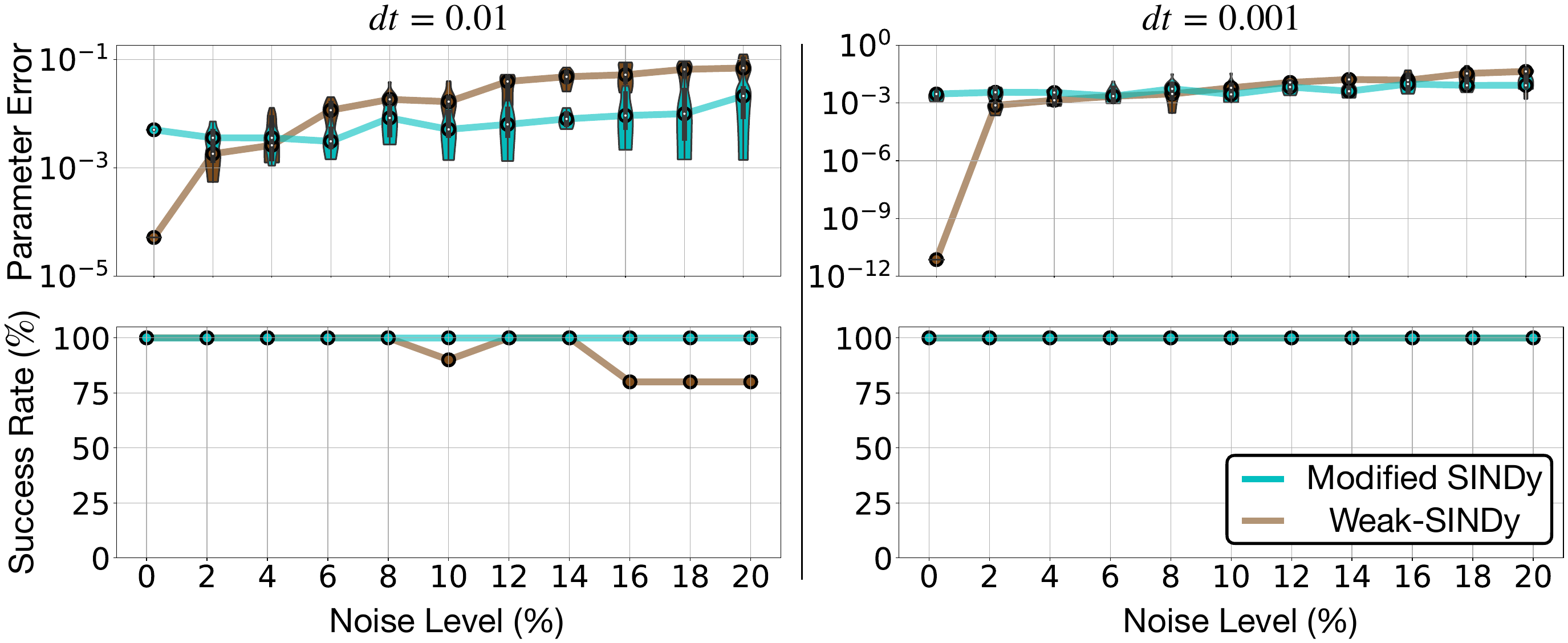}
    \caption{This figure shows noise robustness comparison of modified SINDy and Weak-SINDy using Lorenz attractor as an example. The effect of noise on the parameter error and successful identification rate is compared for both approaches. As it shows in the figure, Weak-SINDy and modified SINDy has almost the same accuracy when $dt=0.001$. However, when $dt=0.01$, the modified SINDy has a slightly higher success identification rate.}
    \label{CompareWeakSINDy}
\end{figure}

This section shows the noise robustness comparison of Weak-SINDy~\cite{messenger2020weaksindy} and modified SINDy using Lorenz attractor as an example. The Lorenz attractor is simulated by setting $x_0=[5,5,25]$, $T=25$, $dt=0.01$ and $dt=0.001$. For both approaches, the library is constructed using up to second order terms~(without constant term). Different percentage of noise is added to the clean data to generate noisy training data. The parameter error and success rate is computed for both approaches. For modified SINDy, Adam optimizer with learning rate of $0.001$ is used to perform the optimization. The sparsity parameter is chosen as $\lambda=0.1$ for most of the time. If the modified SINDy can not produce the correct result, $\lambda=0.15$ is used instead. When $dt=0.01$ we pre-smooth the data using approach mentioned in Sec.~\ref{sec:CompareNoiseNNandModifiedSINDy} and  no pre-smoothing is done when $dt=0.001$. For Weak-SINDy, when $dt=0.01$, $200$ test functions with polynomial order of $14$ are used. The width-at-half-max parameter $r_{whm}=8$, and the support size $s=31$. When $dt=0.001$, $1000$ test functions with polynomial order of $2$ are used. The $r_{whm}=16$, and $s=30$. $30$ different sparsity parameters evenly ranges from $0$ to $0.95$ are used, each generates a different candidate model for Weak-SINDy. The final model we used to calculate the parameter error for Weak-SINDy is the model that has correct structure~(with only correct terms are selected from the library). If the Weak-SINDy fails to produce the model with correct active terms, the model that predicts the test data best is used to calculate the prediction error, and the test data is generated using initial condition $x_{0,test}=[-10,10,15]$ and simulated with $T=25$ and $dt=0.01$. The final comparison result of the best model generated by Weak-SINDy and modified SINDy can be seen in Fig.~\ref{CompareWeakSINDy}.
\label{Appendix:ExampleDetails}
\begin{table}[h]
\centering
\caption{Parameters used in Fig.~\ref{SumExamples}}
\label{ExmapleParameters}
\begin{tabular}{|c||c|c|c|c|c|c|}
\hline
Models         & Initial Condition & Library Order & Learning Rate & $T$  & $q$ & $\lambda$    \\ \hline\hline
Van der Pol    & $[-2,1]$          & $3$           & $0.001$       & $10$ & $1$ & $0.05$       \\ \hline
Duffing        & $[-2,-2]$         & $3$           & $0.001$       & $25$ & $1$ & $0.05$       \\ \hline
Cubic          & $[0,2]$           & $3$           & $0.001$       & $25$ & $1$ & $0.08$       \\ \hline
Lotka-Volterra & $[1,2]$           & $3$           & $0.001$       & $10$ & $1$ & $0.2$        \\ \hline
Lorenz         & $[5,5,25]$        & $2$           & $0.001$       & $25$ & $3$ & $0.1$,$0.15$ \\ \hline
\end{tabular}%
\end{table}
\section{Parameters Used in Fig.~\ref{SumExamples}}
In this section, the models used to simulate the system in Fig.~\ref{SumExamples} are listed. The model used for simulating the Duffing oscillator is
\begin{equation}
    \label{eq:duffing}
    \begin{aligned}
        \dot{x}&=y,\\
        \dot{y}&=-p_1y-p_2x-p_3x^3,
    \end{aligned}
\end{equation}
with $p_1=0.2$, $p_2=0.1$, and $p_3=1$. The model used for simulating the Cubic oscillator is 
\begin{equation}
    \label{eq:cubic}
    \begin{aligned}
        \dot{x}&=p_1x^3+p_2y^3,\\
        \dot{y}&=p_3x^3+p_4y^3,
    \end{aligned}
\end{equation}
with $p_1=-0.1$, $p_2=2$, $p_3=-2$, and $p_4=0.1$. The model used for simulating the Lotka-Volterra system is
\begin{equation}
    \label{eq:lotka}
    \begin{aligned}
        \dot{x}&=p_1x-p_2xy,\\
        \dot{y}&=p_2xy-2p_1y,
    \end{aligned}
\end{equation}
with $p_1=1$ and $p_2=0.5$. Other parameters used for training the modified SINDy is summarized in Table.~\ref{ExmapleParameters}. For all examples, $N_{loop}=5$ and $dt=0.01$.

\section{Tips on Learning Non-Zero Mean Noise}
\begin{figure}[t]
    \centering
    \includegraphics[width=1\textwidth]{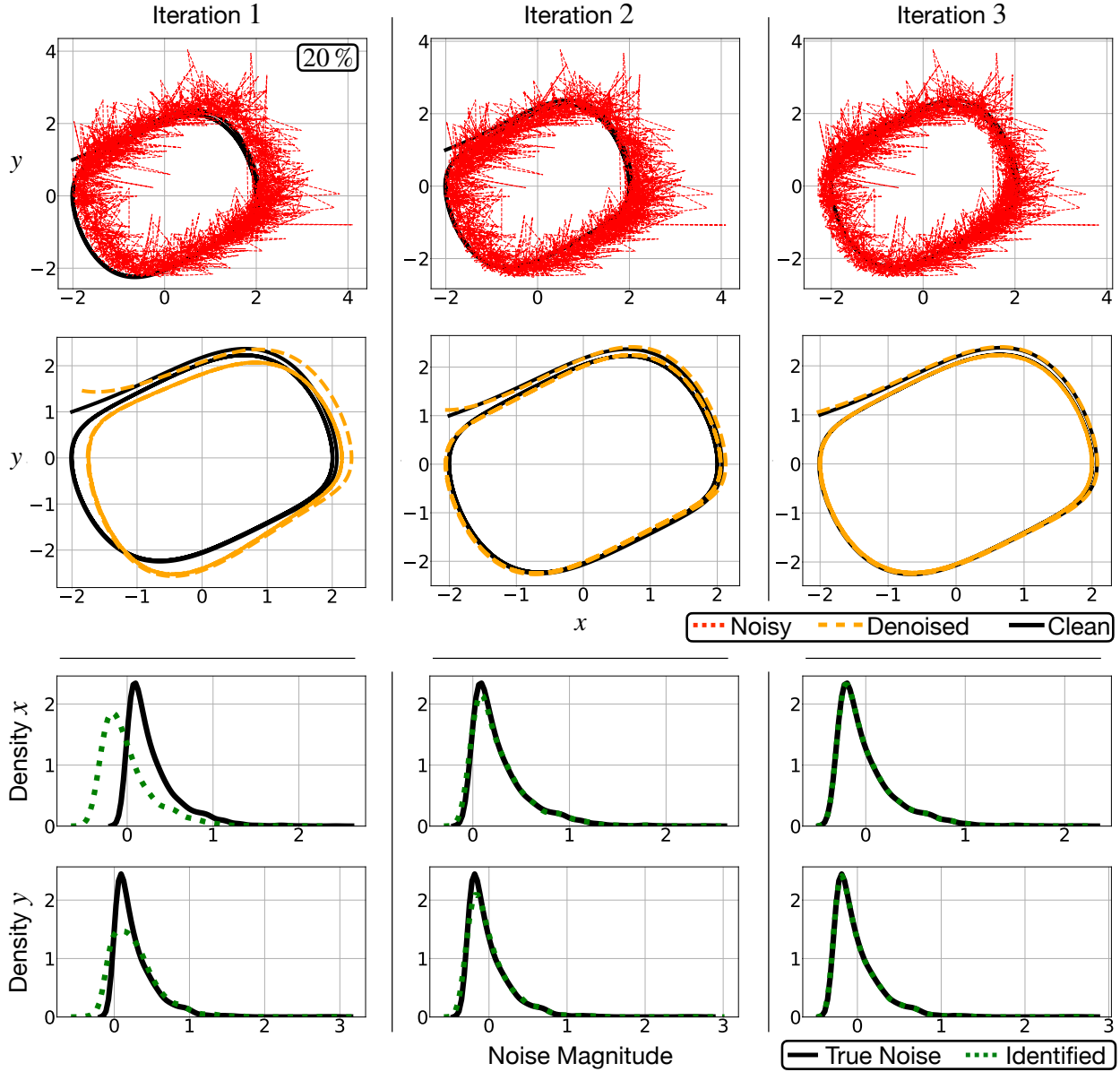}
    \caption{Iterative learning process is shown to tackle the non-zero mean noise distribution. By using this iterative process, modified SINDy can tackle noise distribution with non-zero means.}
    \label{FigIterNoneZeroMean}
\end{figure}
\label{Appendix:NonZeroNoise}
As Sec.~\ref{sec:DiffNoise} suggests, learning non-zero mean noise distribution is much harder than learning the zero-mean noise distribution. To achieve better performance on the non-zero mean noise distribution, we propose an iterative learning approach. This approach can be summarized as follow: 1. Apply modified SINDy to the noisy data, briefly learn the distribution of noise. 2. Subtract the mean of learned noise from the noisy measurement, and use the new data to perform the learning. 3. Repeat the step 2 until the result converges and the correct model is found. The Van der Pol oscillator is used to illustrate this approach, and the clean data is generated the same way in Sec.~\ref{sec:vanderpol}. $20\%$ of Gamma noise is added to create the noisy data. The parameters are set as $q=2$, and $\lambda=0.15$. Fig.~\ref{FigIterNoneZeroMean} demonstrate this approach. However, there's no guarantee that this approach will work when the bias of noise is too large, learning the non-zero mean noise is quite hard and careful tuning is needed. We find out using the soft start approach will also help the denoising of non-zero mean noise.

\section{Identifying Noise Distribution Type}
\label{Appendix:NoiseType}
\begin{table}[]
\centering
\caption{The identified noise distribution versus the true distribution. For Gamma distribution, the parameter $k$ represents shape and $\theta$ represents the scale.}
\label{Table:IDNoiseType}
\resizebox{\textwidth}{!}{
\begin{tabular}{|c|c||c|c|c|}
\hline
\begin{tabular}[c]{@{}c@{}}True\\ Distribution\end{tabular} & State & \begin{tabular}[c]{@{}c@{}}True \\ Parameter\end{tabular}                                            & \begin{tabular}[c]{@{}c@{}}Identified \\ Distribution\end{tabular} & \begin{tabular}[c]{@{}c@{}}Identified \\ Parameters\end{tabular}                                                            \\ \hline\hline
\multirow{2}{*}{Gaussian}                                   & $x$   & $\mu=0,\sigma=0.1413$                                                                                & Gaussian                                                           & $\hat{\mu}=0.003,\hat{\sigma}=0.1451$                                                                                       \\ \cline{2-5} 
                                                            & $y$   & $\mu=0,\sigma=0.1439$                                                                                & Gaussian                                                           & $\hat{\mu}=0.009,\hat{\sigma}=0.1439$                                                                                       \\ \hline
\multirow{2}{*}{Uniform}                                    & $x$   & $\mu=0,\sigma=0.1413$                                                                                & Uniform                                                            & $\hat{\mu}=-0.0717,\hat{\sigma}=0.1438$                                                                                     \\ \cline{2-5} 
                                                            & $y$   & $\mu=0,\sigma=0.1439$                                                                                & Uniform                                                            & $\hat{\mu}=-0.0729,\hat{\sigma}=0.1466$                                                                                     \\ \hline
\multirow{2}{*}{Gamma}                                      & $x$   & \begin{tabular}[c]{@{}c@{}}$k=1,\text{loc}=0,\theta=0.1413$\\ $\mu=0.1413,\sigma=0.02$\end{tabular}  & Gamma                                                              & \begin{tabular}[c]{@{}c@{}}$k=3.2714,\text{loc}=-0.095,\theta=0.0722$\\ $\hat{\mu}=0.1409,\hat{\sigma}=0.0211$\end{tabular} \\ \cline{2-5} 
                                                            & $y$   & \begin{tabular}[c]{@{}c@{}}$k=1,\text{loc}=0,\theta=0.1439$\\ $\mu=0.1439,\sigma=0.021$\end{tabular} & Gamma                                                              & \begin{tabular}[c]{@{}c@{}}$k=10.49,\text{loc}=-0.3105,\theta=0.0432$\\ $\hat{\mu}=0.1419,\hat{\sigma}=0.0217$\end{tabular} \\ \hline
\multirow{2}{*}{Dweibull}                                   & $x$   & \begin{tabular}[c]{@{}c@{}}$c=2.07,\text{loc}=0,$\\ $\text{scale}=0.1413$\end{tabular}                 & Dweibull                                                           & \begin{tabular}[c]{@{}c@{}}$\hat{c}=2.064,\text{loc}=0.8\times10^{-5}$,\\ $\text{scale}=0.1408$\end{tabular}                  \\ \cline{2-5} 
                                                            & $y$   & \begin{tabular}[c]{@{}c@{}}$c=2.07,\text{loc}=0,$\\ $\text{scale}=0.1439$\end{tabular}                 & Dweibull                                                           & \begin{tabular}[c]{@{}c@{}}$\hat{c}=2.048,\text{loc}=-2.8\times10^{-5},$\\ $\text{scale}=0.1438$\end{tabular}                 \\ \hline
\multirow{2}{*}{Rayleigh}                                   & $x$   & $\mu=0.1775,\sigma=0.0085$                                                                           & Rayleigh                                                           & $\hat{\mu}=0.1775,\hat{\sigma}=0.0085$                                                                                      \\ \cline{2-5} 
                                                            & $y$   & $\mu=0.1779,\sigma=0.0086$                                                                           & Rayleigh                                                           & $\hat{\mu}=0.1779,\hat{\sigma}=0.0086$                                                                                      \\ \hline
\end{tabular}}
\end{table}
When the noise is identified, it might be interesting to learn what type of distribution the noise follows. To illustrate this, the Van der Pol oscillator shown in Eq.~\eqref{eq:VanderPol} is simulated with initial condition $[-2,1]$, $T=50$, and $dt=0.001$ (for Gamma and Rayleigh noise distribution, $dt=0.01$). Next, $10\%$ of noise is added to the simulation data to generate the noisy data. The noisy data is provided to modified SINDy to learn the dynamics and identify the noise added to the signal. We set $q=2$, $\lambda=0.15$ ($\lambda=0.2$ for Gamma noise). Adam optimizer with learning rate equals to $0.001$ is used, and the library order is set to $3$. For all cases, the modified SINDy correctly identified the model. As Table.~\ref{Table:IDNoiseType} shows, five different noise distributions is used to generate the noisy data. After the noise is identified, the distribution of noise is fitted into seven candidate noise distributions, which are normal distribution, uniform distribution, Gamma distribution, Dweibull distribution, Rayleigh distribution, Cauchy distribution, and Beta distribution. Next, the sum of the square errors between the $\bhN$ and the fitted distribution is calculated, and the distribution that produces the lowest error is selected as the identified noise distribution. Notice that when there's not enough data provided, it is totally possible that other kinds of distribution is misidentified as the true underlying distribution of noise. The study of how many data points is needed to identify the correct distribution is beyond the scope of this paper. The final result can be summarized in Table.~\ref{Table:IDNoiseType}.

\section{Caveats of the Approach}
\label{sec:Caveats}
This section provides some tips on using modified SINDy.
\begin{enumerate}
    \item Properly design the library: Building the correct library for the regression is the most important part of this algorithm. If the library does not contain the terms included in the actual dynamics, the algorithm will fail to produce the correct noise and system model. Thus, whenever possible, one should include any prior information of the dynamics to build the library. In general, the library needs to be large enough to include all the terms that show up in the dynamics, and at the same time small enough to ensure the robustness. Do not expect the modified SINDy will work on a library with hundreds or thousands of terms, it will break if the library is too large. For example, when using the Lorenz example with above $20\%$ noise, the maximum order of the library modified SINDy can handle is $4$ (about $32$ terms). This happens since the higher order terms in the library will tend to mess up the forward and backward simulation and producing the \texttt{nan} cost, making the optimizer fails. To leverage this, one can try to decrease the learning rate of the optimizer, pre-smooth the data, get better initial estimate of $\bXi$, reduce the library size, or set optimization parameters type as \texttt{float64}. Moreover, whether the constant term $\mathbf{1}$ should be included is case-specific. If the actual dynamics do not have a constant term and the measurement noise is non-zero mean or has significant outliers, including the constant basis in the library will make modified SINDy get stuck at the local minimum more easily. It is advised that the user tries both the library with and without constant basis. 
    \item Initial guess of $\bhN$ and $\bXi$: Having a good initial guess of the estimated noise $\bhN$ and estimated selection parameter $\bXi$ can improve the condition of the optimization problem and allowing us tackle harder problem with more library terms. If possible, the initial values of $\bhN$ can be obtained by pre-smoothing the noisy signal, which will provide a good start for the optimization problem, and it is also good for estimating $\bXi$. If no other information is given, the initial guess of the $\bhN$ can be set as zeros.
\end{enumerate}

\end{document}